\documentclass[aps,pre,reprint,amsmath,amssymb,superscriptaddress]{revtex4-2}

\usepackage{graphicx}
\usepackage{bm}
\usepackage{xcolor}
\usepackage[colorlinks=true,linkcolor=blue,citecolor=blue,urlcolor=blue]{hyperref}

\newcommand{\Pmap}{\mathcal{P}}
\newcommand{\lmax}{\lambda_{\max}}
\newcommand{\vz}{\mathbf{z}}

\begin{document}

\title{Banded Chaos and Time-Semicrystalline Order}

\author{Matteo Baggioli}
\email{b.matteo@sjtu.edu.cn}
\affiliation{School of Physics and Astronomy, Shanghai Jiao Tong University, Shanghai 200240, China}
\affiliation{Wilczek Quantum Center, School of Physics and Astronomy, Shanghai Jiao Tong University, Shanghai 200240, China}

\begin{abstract}
Recent work has introduced the notion of a time semicrystal, a driven nonequilibrium regime in which subharmonic temporal order survives inside deterministic chaos. Here we revisit this phenomenology from the viewpoint of classical nonlinear dynamics. Banded chaotic attractors consist of a finite number of chaotic sectors cyclically permuted by stroboscopic evolution, allowing discrete subharmonic spectral components to coexist with a broadband chaotic continuum. Exemplified through the driven Duffing oscillator and the logistic map, this structure naturally reproduces the characteristic signatures of time-semicrystalline dynamics. Within this framework, periodic subharmonic motion can enter chaos through the familiar period-doubling cascade while retaining a well-defined temporal skeleton governed by the universal Feigenbaum scaling. Temporal order within chaos can also reorganize or collapse through distinct mechanisms, producing continuous or abrupt transitions that may be intertwined with symmetry changes. These results connect time-semicrystal phenomenology with established structures of dissipative nonlinear dynamics, showing that its essential ingredients already arise in standard nonlinear systems and providing a simple framework for extending and testing the paradigm in more complex, many-body, and experimental settings.
\end{abstract}

\maketitle

% =====================================================================
\section{Introduction}
% =====================================================================

Order and chaos are commonly regarded as opposite limits of dynamical behavior. Deterministic chaos~\cite{Ott2002} is characterized by an exponential sensitivity to initial conditions---the celebrated butterfly effect---and appears across an enormous range of nonlinear systems, from atmospheric dynamics~\cite{Lorenz1963} and population models~\cite{May1976} to driven mechanical systems and turbulent flows~\cite{Strogatz2015}. Although the equations of motion are fully deterministic, individual trajectories become unpredictable at long times. Periodicity, by contrast, represents one of the simplest forms of dynamical order: the system repeatedly returns to the same sequence of states. This apparent opposition raises a basic question: can chaotic dynamics retain a form of temporal order?

Classical nonlinear dynamics has long suggested that it can. Chaotic attractors may retain remnants of periodic organization, as in noisy periodicity and in coherent spectral peaks superimposed on broadband continua~\cite{Lorenz1980,Farmer1980,Crutchfield1982}, and chaotic oscillators can lock their phases or frequencies while their amplitudes remain irregular~\cite{Rosenblum1996}. Chaos and temporal organization are therefore not mutually exclusive; the question is rather which forms of temporal order survive in chaotic regimes, and how they are eventually lost.

A particularly striking realization of temporal order is provided by time crystals~\cite{Wilczek2012}. In analogy with ordinary crystals, which spontaneously organize in space, time crystals spontaneously organize in time~\cite{Zaletel2023}. While time-translation symmetry cannot be broken by conventional ground states under standard equilibrium assumptions~\cite{Bruno2013,Watanabe2015} (see, however, Ref.~\cite{Niemi2025}), discrete time crystals (DTCs)~\cite{Else2020} can emerge in periodically driven nonequilibrium systems. In this case, observables develop a robust subharmonic response with a period that is an integer multiple of the driving period $T_d$, thereby spontaneously breaking the discrete time-translation symmetry imposed by the drive~\cite{Khemani2016,Else2016,Yao2017}.

Time-crystalline behavior has been observed in a variety of quantum many-body platforms, from trapped ions and solid-state spins to driven Rydberg gases, e.g., Refs.~\cite{Zhang2017,Choi2017,Liu2025}, and higher-order (period-$n$) responses have been proposed~\cite{Pizzi2019}. More recently, the same underlying idea has been extended beyond the quantum domain. Time-translation symmetry breaking has been proposed in purely classical settings~\cite{Shapere2012}. Classical time-crystalline order has since been analyzed in noisy many-body systems~\cite{Yao2020}, prethermal classical spin systems~\cite{Pizzi2021}, classical Hamiltonian systems~\cite{Khasseh2019}, coupled parametrically driven oscillators~\cite{Heugel2019}, and stochastic many-body models~\cite{Gambetta2019}. Classical time- and spacetime-crystalline states have now been reported experimentally in nematic and chiral liquid crystals~\cite{Zhao2025,Zhao2026}, active granular matter~\cite{Liu2026}, acoustically levitated particles with nonreciprocal interactions~\cite{Morrell2026}, and nanoelectromechanical systems~\cite{Sarkar2026}. Time-crystalline order should therefore be viewed more broadly as a form of nonequilibrium temporal organization rather than as an exclusively quantum phenomenon.

An equally fundamental question concerns how such temporal order is lost. Previous studies have explored the melting of time-crystalline states under thermal fluctuations, changes in density, and other control parameters~\cite{Liu2026,Yue2022,Greilich2024}. The role of deterministic chaos has attracted growing attention: chaos has been discussed in semiclassical and classical models of time crystals~\cite{Khasseh2019,Russomanno2017}, experiments have followed driven time crystals from synchronized motion into chaotic dynamics~\cite{Greilich2024,Greilich2025}, other work has identified distinct routes out of time-crystalline states, such as torus bifurcations~\cite{Cosme2025}, and a symmetry-broken chaotic phase, termed a time glass, has been proposed~\cite{Haga2026}. The issue is conceptually subtle: time-crystalline dynamics is defined by the persistence of a sharp subharmonic temporal structure, whereas chaos is conventionally associated with the loss of long-time predictability. Must the onset of chaos destroy time-crystalline order, or can the two coexist?

A recent answer has come from holography, or gauge--gravity duality~\cite{Ammon2015}, which has become a versatile framework for strongly coupled matter far from equilibrium~\cite{Liu2020}. Within this framework, driven holographic superfluids have been shown to develop spontaneous spatiotemporal order, forming dissipative spacetime supersolids~\cite{Yang2023}. Studying the melting of a discrete time crystal in a periodically driven holographic model with spontaneous $\mathbb{Z}_2$ symmetry breaking, Ref.~\cite{Lei2026} found that, upon tuning the driving frequency, the loss of temporal order proceeds through two distinct stages. Before the subharmonic response disappears completely, an intermediate regime emerges in which sharp subharmonic peaks coexist with a positive largest Lyapunov exponent and a broadband spectral continuum. This regime was termed a \emph{time semicrystal}; its onset displays critical scaling of the Lyapunov exponent, and transitions between distinct periodic skeletons show log-periodic corrections to power-law scaling~\cite{Lei2026}. Since the large-$N$ holographic description is governed by deterministic and dissipative gravitational dynamics, these findings raise a natural question: which aspects of the time-semicrystal phenomenology are specific to holography, and which follow from generic properties of dissipative dynamical systems? Answering it requires identifying the elementary dynamical mechanism that allows temporal order and chaos to coexist, and assessing to what extent the two-stage scenario is generic.

Here we show that this phenomenology has a natural interpretation in classical nonlinear dynamics in terms of \emph{banded chaos}: a chaotic attractor that decomposes into a finite number of bands cyclically permuted by the stroboscopic evolution~\cite{Brown1986,Provatas1991}. The band index then retains an exact cyclic order, which produces sharp subharmonic spectral lines, while the chaotic motion within each band generates a broadband continuum (Fig.~\ref{fig:1}).

Chaotic attractors with cyclic and spectral structure of this kind have a long history in nonlinear dynamics. Their band structure, periodic component, and power spectra were analyzed already in the late 1970s and 1980s~\cite{Lorenz1980,Farmer1980,Crutchfield1982,Brown1986,Provatas1991,Grossmann1977}; early studies of the driven Duffing oscillator documented transitions from subharmonic periodic motion to structured chaotic attractors and analyzed their stroboscopic geometry and power spectra~\cite{Ueda1979}; and the transitions that create, reorganize, and destroy banded attractors, namely period-doubling cascades, band mergings, and crises, are well understood~\cite{Feigenbaum1978,HubermanRudnick1980,GrebogiOttYorke1982,GrebogiOttYorke1983}. Viewed through the language of time-crystalline order, this classical structure provides a natural interpretation of the time-semicrystal phenomenology: the cyclic organization of the bands supplies a discrete temporal skeleton even though the motion within each band is chaotic.

In making this connection, it is useful to distinguish a time-crystalline \emph{phase} from the low-dimensional dynamical analogues considered below. A stable subharmonic orbit of a single nonlinear oscillator does not by itself constitute a thermodynamic phase of matter: it may coexist with other attractors, including the harmonic response, and it acquires a finite switching lifetime in the presence of noise~\cite{Yao2020,Zhang2017PRA}. We therefore use low-dimensional models as minimal realizations of the relevant temporal symmetry structure. In this operational sense, a periodic subharmonic state is the deterministic analogue of discrete time-crystalline order, and a banded-chaotic state the analogue of time-semicrystalline order; we use the labels DTC and TSC regimes in this sense throughout. The question of genuine many-body robustness lies beyond the scope of the present work.

The purpose of this work is to use the Duffing oscillator and the logistic map as controlled benchmark systems for making this connection quantitative. These models allow the periodic, banded-chaotic, and single-band chaotic regimes to be identified unambiguously and related to standard bifurcations of dissipative nonlinear dynamics. We use them to formulate operational diagnostics of the surviving cyclic order and to examine how this order is created, reorganized, and ultimately lost. In this way, the recent time-semicrystal phenomenology can be placed within the broader framework of classical nonlinear dynamics while keeping separate the additional question of robustness required for a genuine many-body time-crystalline phase.

% =====================================================================
\section{Banded chaos and time semicrystals}
\label{sec:banded}
% =====================================================================

Let us consider a single-degree-of-freedom classical system subject to an external periodic drive of period $T_d=2\pi/\omega_d$. Owing to the periodicity of the forcing, the long-time dynamics can be conveniently characterized by the stroboscopic map $\Pmap$, obtained by sampling the trajectory once every driving period. Defining the stroboscopic state as
\begin{equation}
\vz_n \equiv (x_n,v_n) = \left[x(nT_d),v(nT_d)\right], \qquad \vz_{n+1}=\Pmap(\vz_n),
\label{eq:strobe}
\end{equation}
where $v(t)\equiv\dot x(t)$, the continuous-time dynamics is reduced to the iteration of a discrete map. The same language applies directly to an intrinsically discrete system, $x_{n+1}=f(x_n)$, for which $x_n$ itself is the state variable. 

If the asymptotic response has the same period as the external drive, $x(t+T_d)=x(t)$, all stroboscopic samples coincide and the attractor of $\Pmap$ consists of a single point. More generally, a response with period $pT_d$ generates a $p$-cycle,
\begin{equation}
\vz_0 \longrightarrow \vz_1 \longrightarrow \cdots \longrightarrow \vz_{p-1} \longrightarrow \vz_0 .
\label{eq:pcycle}
\end{equation}
A period-$p$ subharmonic response is therefore represented by a $p$-cycle of $\Pmap$ and contains a coherent response at $\omega_d/p$. In the present low-dimensional setting, this provides the simplest deterministic realization of discrete time-crystalline order.

\begin{figure*}[t]
\centering
\includegraphics[width=\textwidth]{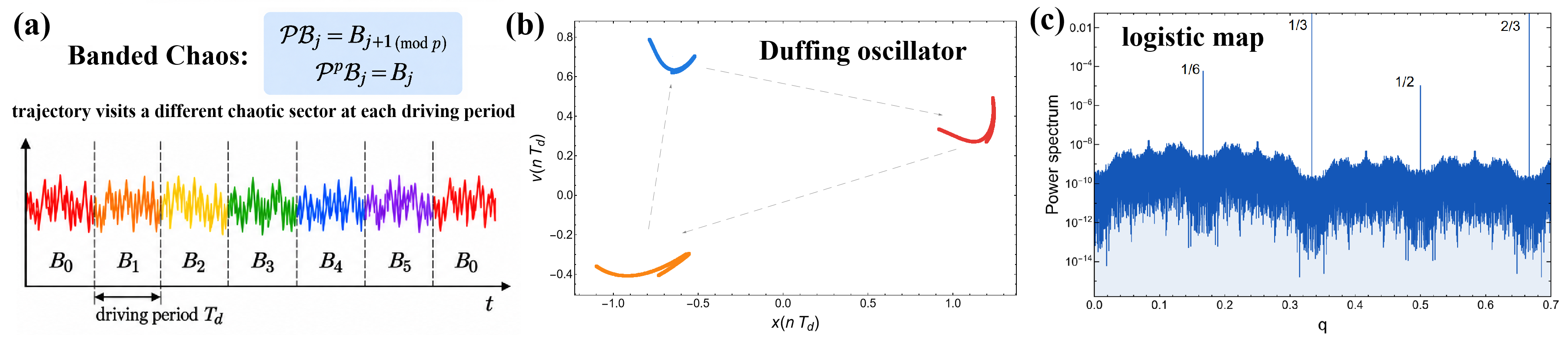}
\caption{\textbf{Banded chaos as temporal order within chaos.} \textbf{(a)}~Schematic representation of a $p$-band chaotic attractor. The stroboscopic map $\Pmap$ cyclically permutes the chaotic bands, $\Pmap B_j=B_{j+1\,(\mathrm{mod}\,p)}$, so that a trajectory returns to the same band only after $p$ driving periods, $\Pmap^p B_j=B_j$. Despite the irregular dynamics within each band, the band index retains an exact finite temporal order. \textbf{(b)}~Stroboscopic Poincar\'e section of the driven Duffing oscillator for $\alpha=-1$, $\beta=1$, $\gamma=0.30$, $F=0.58$, and $\omega_d=1.36$, sampled once per driving period $T_d=2\pi/\omega_d$. The three colors identify the three cyclically visited chaotic bands. The trajectory was initialized at $(x_0,\dot x_0)=(1,0)$; transients were discarded. \textbf{(c)}~Power spectrum of the logistic map for $r=3.850975$, in the six-band chaotic regime. Discrete components at reduced frequencies $q=1/6$, $1/3$, $1/2$ and $2/3$ coexist with a broadband chaotic continuum. Panels (b) and (c) illustrate different representative band numbers, emphasizing that the finite cyclic structure is not tied to a specific model or value of $p$.}
\label{fig:1}
\end{figure*}

A richer situation arises in \emph{banded chaos}~\cite{Brown1986,Provatas1991}. In this case, the $p$ points of the periodic orbit broaden into $p$ extended chaotic subsets while their cyclic permutation under one driving period survives [Fig.~\ref{fig:1}(a)]. The chaotic stroboscopic attractor can then be decomposed as
\begin{equation}
\mathcal{A} = \bigcup_{j=0}^{p-1} B_j ,
\label{eq:bands}
\end{equation}
with
\begin{equation}
\Pmap(B_j) = B_{j+1\,(\mathrm{mod}\,p)} .
\label{eq:perm}
\end{equation}
Thus each driving period transfers the trajectory from one chaotic sector to the next,
\begin{equation}
B_0 \xrightarrow{\ \Pmap\ } B_1 \xrightarrow{\ \Pmap\ } \cdots \xrightarrow{\ \Pmap\ } B_{p-1} \xrightarrow{\ \Pmap\ } B_0 ,
\label{eq:sequence}
\end{equation}
whereas only the $p$-fold map leaves each individual band invariant,
\begin{equation}
\Pmap^{p}(B_j) = B_j .
\label{eq:pfold}
\end{equation}
The dynamics generated by $\Pmap^p$ within a given band may itself be chaotic even though the permutation of the bands remains exactly periodic. Banded chaos therefore separates two distinct structures: unpredictable motion within each band and a finite cyclic organization among the bands. It is this surviving cyclic organization that we identify with the temporal skeleton of a time-semicrystalline state.

The coexistence of chaos and temporal order is particularly transparent in the spectrum. Let
\begin{equation}
O_n \equiv O(\vz_n)
\end{equation}
be an observable sampled stroboscopically, and denote by
\begin{equation}
a_j \equiv \langle O\rangle_{B_j}
\end{equation}
its conditional average over the $j$th chaotic band. Because the bands are visited cyclically, the observable can be written schematically as
\begin{equation}
O_n = a_{n\,(\mathrm{mod}\,p)} + \delta O_n ,
\label{eq:decomp}
\end{equation}
where the first contribution is exactly $p$-periodic and $\delta O_n$ contains the chaotic fluctuations within the bands.

We define the coherent component of the stroboscopic sequence at reduced
frequency $q\in[0,1)$, corresponding to a stroboscopic angular frequency
$q\omega_d$ defined modulo $\omega_d$, as the part of the response that
survives averaging over long times:
\begin{equation}
\Phi_q[O] \equiv \lim_{N\to\infty}\frac{1}{N}\sum_{n=0}^{N-1}
O_n\, e^{-2\pi i q n}.
\label{eq:Phi}
\end{equation}
Thus, $\Phi_q[O]$ measures the strength of the persistent Fourier component
of the stroboscopic response at reduced frequency $q$. For a $p$-cyclic
state, the coherent reduced frequencies are
\begin{equation}
q=\frac{m}{p},\qquad m=0,\dots,p-1,
\end{equation}
and the corresponding amplitudes are
\begin{equation}
\Phi_{m/p} = \frac{1}{p}\sum_{j=0}^{p-1} a_j\, e^{-2\pi i m j/p}.
\label{eq:Phimp}
\end{equation}
We denote the corresponding coherent amplitude and spectral weight by
\begin{equation}
A_{m/p}\equiv|\Phi_{m/p}|, \qquad W_{m/p}\equiv|\Phi_{m/p}|^2 .
\label{eq:AW}
\end{equation}
The corresponding stroboscopic spectral measure is defined from the long-time autocorrelation function
\begin{equation}
C(k)\equiv \lim_{N\to\infty}\frac{1}{N}
\sum_{n=0}^{N-1} O_{n+k}O_n^{\ast},
\end{equation}
through its Fourier transform,
\begin{equation}
S(q)\equiv \sum_{k=-\infty}^{\infty} C(k)e^{-2\pi i q k},
\qquad q\in[0,1),
\end{equation}
where $S(q)$ is understood in the sense of a spectral measure. For a $p$-cyclic chaotic state, this measure naturally decomposes into a pure-point contribution associated with the cyclic permutation of the bands and a continuous contribution generated by chaotic fluctuations within them,
\begin{equation}\label{eq:Sq}
S(q)=\sum_{m=0}^{p-1}
W_{m/p}\,
\delta\!\left(q-\frac{m}{p}\right)
+S_{\rm cont}(q),
\qquad q\in[0,1).
\end{equation}
The discrete weights are $W_{m/p}=|\Phi_{m/p}|^2$, with $\Phi_{m/p}$ given by Eq.~\eqref{eq:Phimp}.
A step-by-step derivation of Eq.~\eqref{eq:Sq} is given in the Supplemental Material.
Here the discrete contribution reflects the cyclic permutation of the bands, whereas $S_{\rm cont}(q)$ arises from chaotic fluctuations within them. This decomposition assumes that the chaotic motion within each band does not
itself contain additional persistent periodic components; otherwise, extra
sharp spectral lines may appear.

Equation~\eqref{eq:Sq} describes the spectrum of the stroboscopically sampled
sequence, obtained by observing the system only once per driving period.
The continuous-time trajectory $x(t)$ contains more information and, for a
period-$pT_d$ structure, has spectral lines at integer multiples of
$\omega_d/p$. Different continuous-time harmonics can become indistinguishable
after stroboscopic sampling, and therefore the coefficients $\Phi_{m/p}$
should not be interpreted as the amplitudes of individual lines in the
continuous-time spectrum. For example, under sampling once every $T_d$, the continuous-time components
at $\omega_d/3$ and $4\omega_d/3$ produce the same stroboscopic oscillation.

Within this framework, a time-semicrystalline state is characterized
operationally by the coexistence of
\begin{equation}
\lambda_{\max}>0
\qquad\text{and}\qquad
A_{m/p}>0
\end{equation}
for at least one nontrivial subharmonic $m/p$, with the coherent amplitude
understood in the long-time limit $N\to\infty$. The first condition
identifies chaos, while the second identifies persistent coherent temporal
order. Neither condition alone is sufficient: a periodic subharmonic state
has $A_{m/p}>0$ but is not chaotic, whereas single-band chaos has
$\lambda_{\max}>0$ but no finite cyclic subharmonic structure.

We study this phenomenology using two standard nonlinear systems. The first is the periodically driven Duffing oscillator~\cite{Ueda1979,NayfehMook1979},
\begin{equation}
\ddot x + \gamma\dot x + \alpha x + \beta x^3 = F\cos(\omega_d t),
\label{eq:duffing}
\end{equation}
where $\gamma$ is the damping coefficient, $\alpha$ and $\beta$ determine the restoring potential, and $F$ and $\omega_d$ are the amplitude and frequency of the external drive. We mainly use $\omega_d$ as the control parameter while keeping the remaining coefficients fixed. The Duffing oscillator provides a minimal setting displaying nonlinear resonance, multistability, period doubling, banded chaos, and crisis phenomena.

The second model is the logistic map~\cite{May1976},
\begin{equation}
x_{n+1} = r x_n(1-x_n),
\label{eq:logistic}
\end{equation}
where $x_n$ is a dimensionless variable normalized to $0\leq x_n\leq1$,
with unity setting the saturation scale. The parameter $r$ controls the
nonlinearity of the map and is used as the control parameter. The factor
$(1-x_n)$ suppresses the growth as $x_n$ approaches its upper bound. Changing $r$ generates the familiar hierarchy of periodic windows,
period-doubling cascades, banded-chaotic attractors, and band-merging crises.

Figures~\ref{fig:1}(b)--\ref{fig:1}(c) illustrate the time-semicrystalline structure in the two models. In Fig.~\ref{fig:1}(b), the stroboscopic Poincar\'e section of the Duffing oscillator consists of three chaotic bands visited cyclically by the trajectory. Thus $\Pmap$ is chaotic within the attractor, whereas the band index retains an exact period-three structure. Figure~\ref{fig:1}(c) shows instead the stroboscopic spectrum $S(q)$ of the logistic map in a six-band chaotic regime. Sharp coherent components, including those at $q=1/6$ and $q=1/3$, coexist with a broadband continuum. The two examples emphasize that the cyclic temporal skeleton is neither tied to a particular model nor to a specific value of $p$.

Figure~\ref{fig:2} follows the corresponding evolution in the Duffing oscillator as the driving frequency is varied. The largest Lyapunov exponent $\lmax$, measured per unit time and evaluated from the variational dynamics using the standard Benettin procedure~\cite{Benettin1980}, with details given in the Supplemental Material, distinguishes periodic motion from chaos. At the onset of the intermediate regime, the dynamics becomes chaotic while
the coherent subharmonic response remains finite. The system therefore loses
periodicity without losing its cyclic temporal organization.

The distinction is also visible in the continuous-time spectra and Poincar\'e sections shown in Figs.~\ref{fig:2}(b)--\ref{fig:2}(g). In the periodic regime, the spectrum consists only of discrete lines and the stroboscopic attractor is a finite cycle. In the time-semicrystalline regime, the discrete orbit broadens into cyclically permuted chaotic bands: a broadband continuum develops while sharp subharmonic components survive. At larger driving frequency the separated bands ultimately lose their finite cyclic organization and merge into a single chaotic attractor, at which point the coherent subharmonic structure disappears.

The loss of temporal order therefore occurs in two stages. First, periodic motion becomes chaotic while retaining a finite cyclic skeleton; only later is that skeleton itself destroyed. In this sense, the onset of chaos and the loss of temporal order are distinct dynamical events.

% =====================================================================
\section{Role of the spatiotemporal $\mathbb{Z}_2$ symmetry}
\label{sec:Z2}
% =====================================================================

The bifurcation sequence of the Duffing oscillator is further constrained by a spatiotemporal $\mathbb{Z}_2$ symmetry. More generally, for the even Duffing potential associated with Eq.~\eqref{eq:duffing}, the equation of motion is invariant under
\begin{equation}
\mathcal{S}:\quad x(t)\longrightarrow -x\!\left(t+\frac{T_d}{2}\right).
\label{eq:S}
\end{equation}
This symmetry does not rely specifically on the double-well regime: it follows from the evenness of the potential. Periodic and chaotic attractors may either be invariant under $\mathcal{S}$ or occur in symmetry-related pairs.

\begin{figure}[!t]
\centering
\includegraphics[width=\columnwidth]{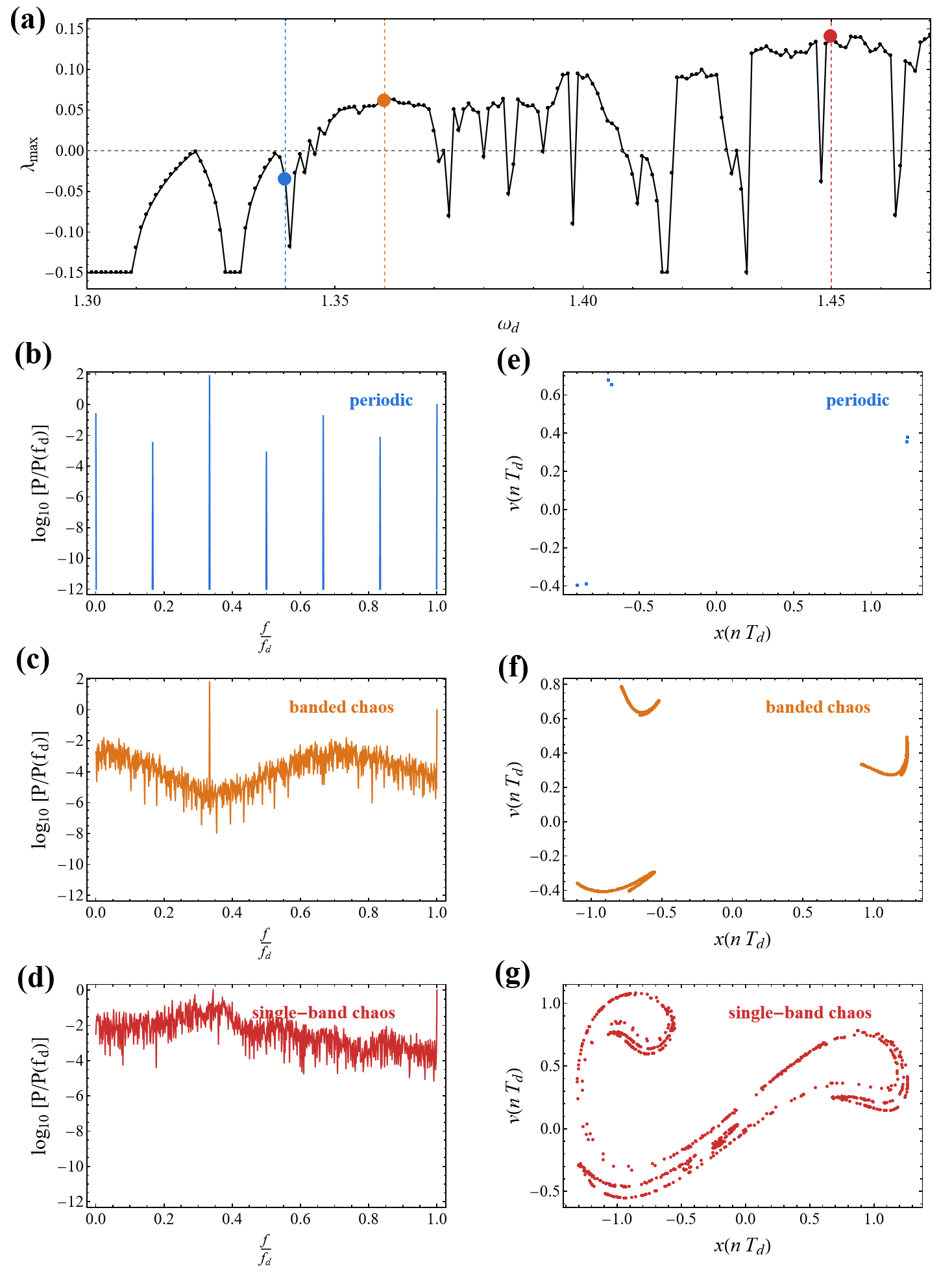}
\caption{
\textbf{Representative periodic, banded-chaotic, and single-band chaotic
dynamics in the driven Duffing oscillator.}
The driving frequency $\omega_d$ is varied at fixed
$\alpha=-1$, $\beta=1$, $\gamma=0.30$, and $F=0.58$, with initial condition
$(x_0,\dot{x}_0)=(1,0)$. \textbf{(a)} Largest Lyapunov exponent $\lambda_{\max}$ of the attractor
reached from the initial condition $(x_0,\dot{x}_0)=(1,0)$, as a function
of $\omega_d$. The dependence on driving frequency is nonmonotonic and contains
periodic windows and re-entrant chaotic intervals. The three colored symbols
mark representative attractors at $\omega_d=1.34$, $1.36$, and $1.45$;
the vertical dashed lines are guides to these examples and do not denote
phase boundaries.
\textbf{(b)--(d)} Corresponding continuous-time power spectra, normalized to
the spectral weight at the drive frequency $f_d$. The state at
$\omega_d=1.34$ is a period-$6T_d$ orbit and therefore contains the
corresponding subharmonic structure. At $\omega_d=1.36$ the dynamics is
chaotic but remains organized into three cyclically visited bands; the
$\mathcal S$-symmetric attractor carries coherent lines only at odd multiples
of $f_d/3$, as required by the selection rule of Sec.~\ref{sec:Z2}. At
$\omega_d=1.45$ the remaining cyclic organization has been lost and the
dynamics is single-band chaotic.
\textbf{(e)--(g)} Corresponding stroboscopic Poincar\'e sections sampled once
per driving period $T_d=2\pi/\omega_d$. Panel \textbf{(e)} shows the six
points of the period-$6T_d$ orbit, panel \textbf{(f)} the three chaotic bands,
and panel \textbf{(g)} the single-band chaotic attractor.
}
\label{fig:2}
\end{figure}

For the branch used in Fig.~\ref{fig:2}, the low-frequency period-three orbit is $\mathcal{S}$-symmetric and first undergoes a symmetry-breaking bifurcation at $\omega_d\simeq1.3218$ before entering the subsequent period-doubling cascade~\cite{SwiftWiesenfeld1984}. The resulting pair of symmetry-related period-three orbits subsequently undergoes the flip sequence
\begin{equation}
3\to6\to12\to24\to\cdots,
\label{eq:cascadeDuffing}
\end{equation}
with the first few bifurcations at $\omega_d\simeq1.33846$, $1.34330$, $1.34442$, and $1.34467$. Their accumulation near $\omega_d\simeq1.34473$ marks the onset of banded-chaotic dynamics for this branch. The corresponding bifurcation sequence, obtained by Newton--Floquet
continuation of the periodic solutions, is summarized in the Supplemental
Material.

The same symmetry imposes a simple selection rule on the continuous-time spectrum. For an $\mathcal{S}$-symmetric period-three state, the coherent part satisfies
\begin{equation}
x\!\left(t+\frac{3T_d}{2}\right) = -x(t).
\label{eq:selection}
\end{equation}
Expanding the coherent response in harmonics of $\omega_d/3$, this relation eliminates all even harmonics: only odd multiples of $\omega_d/3$ can carry coherent weight. In particular, the DC component and the lines at $2\omega_d/3,4\omega_d/3,\dots$ vanish in the symmetric state and may appear only after the $\mathbb{Z}_2$ symmetry is broken. This selection rule is visible directly in the spectra of Fig.~\ref{fig:2};
a derivation is given in the Supplemental Material.

The symmetry organization persists inside the banded-chaotic regime. Depending on $\omega_d$, the three-band attractor may be either $\mathcal{S}$-symmetric or occur as a pair of symmetry-related chaotic attractors. At larger driving frequency the broken-symmetry chaotic branches ultimately merge in an attractor-merging crisis at $1.39398<\omega^\star<1.39399$. At this transition the $\mathbb{Z}_2$ symmetry is restored and the coherent period-three temporal order disappears. We analyze this final loss of time-semicrystalline order in detail in Sec.~\ref{sec:loss}.

% =====================================================================
\section{Birth of temporal order from nonlinear resonance}
\label{sec:birth}
% =====================================================================

Having established that time-semicrystalline dynamics corresponds to banded chaos, we now ask how the underlying temporal order is generated in the first place. In the Duffing oscillator this question can be addressed analytically in the weakly nonlinear regime, where subharmonic responses arise through standard nonlinear resonances~\cite{NayfehMook1979}. We focus on the $1{:}3$ resonance, which provides a simple route to a period-$3T_d$ discrete time-crystalline state.

We consider Eq.~\eqref{eq:duffing} in the single-well regime $\alpha=\omega_0^2>0$, with weak damping and weak nonlinearity, and tune the drive close to $\omega_d\simeq3\omega_0$. We seek a response containing both the directly forced oscillation at $\omega_d$ and an emergent subharmonic component at $\omega_d/3$,
\begin{equation}
x(t) \simeq a(t)\cos\!\left[\frac{\omega_d}{3}t+\theta(t)\right] + b\cos(\omega_d t-\delta).
\label{eq:ansatz}
\end{equation}
To leading order, the amplitude of the directly driven component is $b=F/\sqrt{(\omega_0^2-\omega_d^2)^2+\gamma^2\omega_d^2}$. A standard averaging over the fast oscillations, treating the amplitude
$a(t)$ and phase $\theta(t)$ as slowly varying, then gives for stationary
subharmonic solutions,
\begin{equation}
\left[\omega_0^2-\left(\frac{\omega_d}{3}\right)^2+\frac{3\beta}{2}b^2+\frac{3\beta}{4}a^2\right]^2+\left(\frac{\gamma\omega_d}{3}\right)^2=\left(\frac{3\beta}{4}\right)^2a^2b^2 .
\label{eq:averaging}
\end{equation}
The averaging derivation and the resulting saddle-node condition are given
in the Supplemental Material.

\begin{figure}[t]
\centering
\includegraphics[width=\columnwidth]{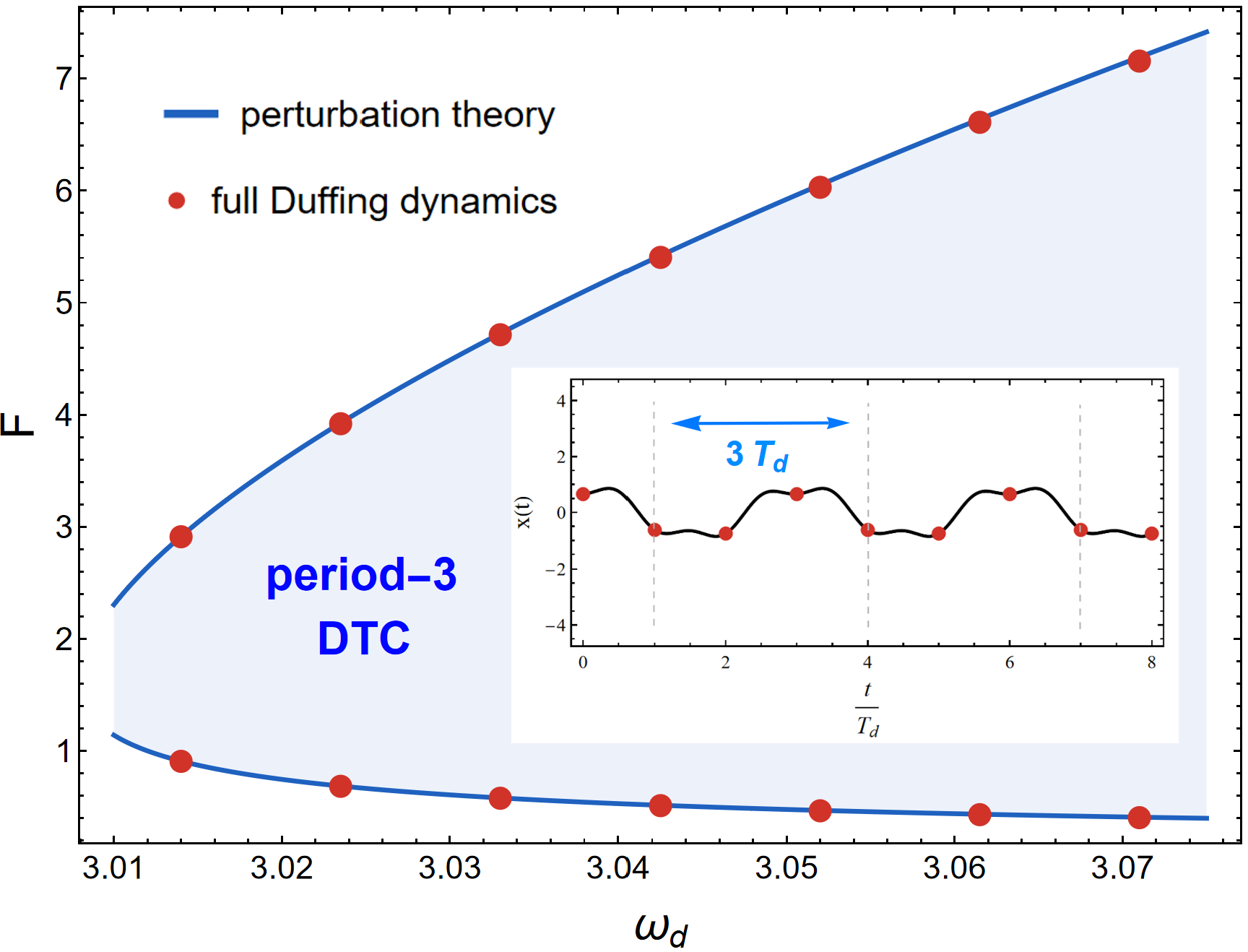}
\caption{\textbf{Birth of period-three temporal order from nonlinear resonance.} The shaded region shows the $1{:}3$ resonance tongue of the Duffing oscillator, i.e., the region where Eq.~\eqref{eq:averaging} admits a real subharmonic amplitude $a\neq0$. Solid blue curves are the perturbative saddle-node boundaries $F_\pm(\omega_d)$ of Eq.~\eqref{eq:tongue}, while red symbols are obtained from the full nonlinear dynamics. Below the lower boundary $F_-$, the drive is too weak for the resonant mixing to overcome damping; above the upper boundary $F_+$, the strong forced response shifts the effective oscillator frequency away from $\omega_d/3$ and the resonance is lost. Inside the tongue, a stable period-three branch coexists with an unstable partner; the two merge at the boundaries, while outside only the harmonic response exists. The inset shows a representative period-$3T_d$ trajectory inside the tongue, with red markers denoting stroboscopic sampling every $T_d$. Throughout, we use $\omega_0=1$, $\gamma=0.002$, and $\beta=0.05$; the inset corresponds to $\omega_d=3.040$ and $F=2$.}
\label{fig:3}
\end{figure}

Equation~\eqref{eq:averaging} predicts a finite region of parameter space where a nonzero subharmonic response exists, namely where Eq.~\eqref{eq:averaging} admits a real solution $a\neq0$. Importantly, the period-three response does not emerge continuously from the
harmonic period-one state. Instead, it appears at finite amplitude through a
saddle-node bifurcation, together with an unstable partner. The period-one
state can therefore remain stable over part of the resonance tongue, leading
to bistability: depending on the initial condition, the system may approach
either the harmonic or the period-three response.

The two saddle-node boundaries of the $1{:}3$ resonance tongue can be written in closed form as
\begin{align}
&F_\pm(\omega_d) = \sqrt{\Delta_1^2+\gamma^2\omega_d^2}\nonumber\\
&\times\left[\frac{8}{21\beta}\left(\left(\frac{\omega_d}{3}\right)^2-\omega_0^2\pm\sqrt{\Delta_2^2-\frac{7\gamma^2\omega_d^2}{9}}\right)\right]^{1/2},
\label{eq:tongue}
\end{align}
where we have defined for brevity $\Delta_1 \equiv \omega_0^2-\omega_d^2$ and $\Delta_2\equiv \omega_0^2-\left(\frac{\omega_d}{3}\right)^2$.

For $\beta>0$, real saddle-node boundaries require $\omega_d>3\omega_0$ and
\begin{equation}
\left[\omega_0^2-\left(\frac{\omega_d}{3}\right)^2\right]^2 \ge \frac{7\gamma^2\omega_d^2}{9}.
\label{eq:tongueexist}
\end{equation}

Figure~\ref{fig:3} provides a direct test of this perturbative prediction. The solid blue curves are the analytical boundaries $F_\pm(\omega_d)$ from Eq.~\eqref{eq:tongue}, while the red symbols are the corresponding saddle-node points obtained from the full Duffing dynamics. Their close agreement demonstrates that the emergence of period-three temporal order is quantitatively captured by the weakly nonlinear theory. The inset shows a representative trajectory inside the resonance tongue: although the system is driven with period $T_d$, the stroboscopic sequence returns to itself only after $3T_d$.

The saddle-node mechanism is not peculiar to the weakly nonlinear example. Continuing the period-three branch used in Fig.~\ref{fig:2} toward lower driving frequency shows that it is likewise created in a saddle-node bifurcation at $\omega_d\simeq1.2821$; the corresponding numerical saddle-node point is reported in the
Supplemental Material. 

The logistic map provides the corresponding discrete-time example as well. Its period-three window is born through a tangent, or saddle-node, bifurcation at $r = 1+\sqrt{8}$, a classic result for the logistic map~\cite{SahaStrogatz1995}. Thus, in both models considered here, period-three temporal order appears through a hard saddle-node birth rather than continuously from the period-one state.

% =====================================================================
\section{From periodic order to banded chaos}
\label{sec:cascade}
% =====================================================================

We now turn to how a periodic time-crystalline state evolves into the time-semicrystalline regime. Starting from a stable period-$3T_d$ response, a natural route
to chaos is the familiar period-doubling cascade, already observed in early studies of subharmonic Duffing dynamics~\cite{Ueda1979}. The important point here is that the successive bifurcations do not immediately destroy temporal organization. Instead, they generate an increasingly fine hierarchy of subharmonic structure while retaining the underlying period-three skeleton.

As $\omega_d$ is varied, the period-three orbit loses stability through a
supercritical period-doubling bifurcation, producing a stable period-six
response. The subsequent bifurcations generate the cascade discussed below;
their numerical locations are reported in Table~\ref{tab:SMfeig} of the
Supplemental Material. We quantify the resulting period-six splitting through
\begin{equation}
D_6 \equiv \frac{1}{3}\sum_{j=0}^{2}\left\|\vz_{j+3}-\vz_j\right\| ,
\label{eq:D6}
\end{equation}
where $\vz_j$ denotes the $j$th stroboscopic state along the period-six orbit, with
$j=0,1,2$, so that $\vz_{j+3}$ is the state reached three driving periods later. Here, $|| \cdot ||$ denotes the Euclidean norm, 
\begin{equation}
    \|\mathbf{z}_{j+3} - \mathbf{z}_j\| = \sqrt{(x_{j+3} - x_j)^2 + (v_{j+3} - v_j)^2},
\end{equation}
with $\mathbf{z} = (x, v)$. Thus $D_6=0$ for a period-three orbit and becomes finite when the two three-period sectors split into a period-six cycle. The corresponding spectral order parameter is $A_{1/6}=|\Phi_{1/6}|$, defined above.

\begin{figure}[!t]
\centering
\includegraphics[width=\columnwidth]{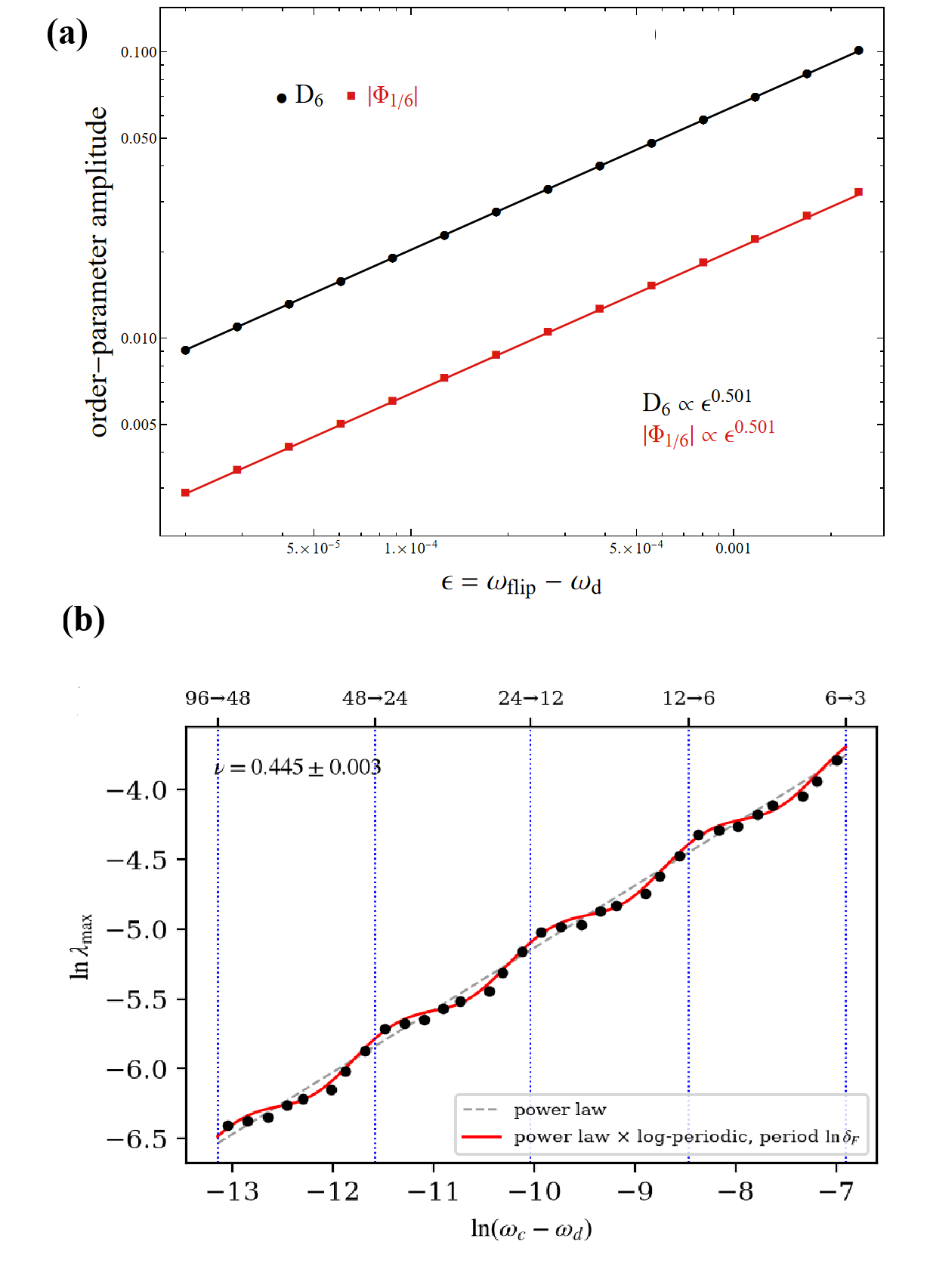}
\caption{
\textbf{From periodic order to banded chaos through the Feigenbaum cascade.}
\textbf{(a)} Scaling at the first flip bifurcation $3\to6$.
The orbit splitting $D_6$ and the emergent subharmonic amplitude
$|\Phi_{1/6}|$ both vanish as $\epsilon^{1/2}$, with fitted exponents
$0.501$, in agreement with the square-root scaling of a supercritical
flip bifurcation.
\textbf{(b)} Lyapunov scaling and log-periodic fine structure on the chaotic
side of the accumulation point, $\omega_c\simeq1.434984$.
Black symbols show $\ln\lambda_{\max}$ obtained from a dense scan of
1000 values of $\omega_d$ and averaged in 32 bins. The gray dashed line is a pure power law, while the red line is
Eq.~\eqref{eq:logperiodic} with the log-period fixed to the Feigenbaum value
$\Delta_{\log}=\ln\delta_F\simeq1.541$, yielding
$\nu=0.445\pm0.003$ (statistical error), with window-dependent values in the
range approximately $0.44$--$0.45$. The leading log-periodic correction
captures the systematic oscillatory fine structure of the data.
Blue dotted lines mark independently determined band-merging points
$\omega_k$ (top labels:
$3\cdot2^k\to3\cdot2^{k-1}$ bands); the steps of the Lyapunov staircase
coincide with these mergers. Their geometric spacing and the corresponding
factor-of-two Lyapunov ratios provide the fit-independent test of
Feigenbaum discrete scale invariance discussed in the text and Supplemental
Material. Parameters are $\alpha=-1$, $\beta=1$, $\gamma=0.2$, and $F=0.74$.
}
\label{fig:4}
\end{figure}

For the quantitative analysis, we use
$\alpha=-1$, $\beta=1$, $\gamma=0.2$, and $F=0.74$, rather than the
parameters of Fig.~\ref{fig:2}. This parameter set was chosen because it
exposes a longer and cleaner sequence of period doublings before the onset
of chaos, allowing the universal scaling to be resolved more accurately.
The cascade develops on an asymmetric period-three branch, as required for
the subsequent flip bifurcations of the driven Duffing oscillator
~\cite{SwiftWiesenfeld1984}. The first flip occurs at
$\omega_{\rm flip}\simeq1.43896$. Defining
$\epsilon=\omega_{\rm flip}-\omega_d$, Fig.~\ref{fig:4}(a) gives
\begin{equation}
D_6\sim\epsilon^{0.501},
\qquad
|\Phi_{1/6}|\sim\epsilon^{0.501},
\end{equation}
in excellent agreement with the square-root scaling expected at the onset of a continuous period-doubling bifurcation. Correspondingly, the spectral weight
$W_{1/6}=|\Phi_{1/6}|^2$ grows linearly in $\epsilon$.

Further period doublings generate the hierarchy $3\to6\to12\to24\to\cdots$, which accumulates at $\omega_c\simeq1.434984$. The onset of chaos can then be characterized independently through the largest Lyapunov exponent. Writing $d=\omega_c-\omega_d>0$ on the chaotic side, Feigenbaum theory predicts the Huberman--Rudnick scaling~\cite{HubermanRudnick1980}
\begin{equation}
\lambda_{\max}\sim d^{\nu_{\rm HR}},
\qquad
\nu_{\rm HR}=\frac{\ln 2}{\ln\delta_F}\simeq0.4498 ,
\label{ruru}
\end{equation}
where $\delta_F\simeq4.6692$ is the Feigenbaum constant~\cite{Feigenbaum1978}. 

The Feigenbaum hierarchy implies the discrete scaling relation
\begin{equation}
\lmax\!\left(\frac{d}{\delta_F}\right)\simeq\frac{1}{2}\,\lmax(d),
\label{eq:discreteRG}
\end{equation}
where $d=\omega_c-\omega_d$ measures the distance from the accumulation point on the chaotic side. This relation can be understood directly from the self-similarity of the cascade. On the chaotic side, the band-merging points accumulate geometrically, $d_k/d_{k+1}\to\delta_F$, so that dividing $d$ by $\delta_F$ moves the system one level deeper into the hierarchy, where the attractor has twice as many bands. Feigenbaum renormalization implies that the dynamics at this deeper level, observed every two driving periods and rescaled in phase space, reproduces asymptotically the dynamics at the level above~\cite{Feigenbaum1978,HubermanRudnick1980}. The trajectory therefore accumulates the same chaotic stretching per return to a given band, but each return takes twice as long. Since a Lyapunov exponent is unaffected by a rescaling of phase-space coordinates and doubles when two iterations are counted as one, the stretching rate is halved, which is Eq.~\eqref{eq:discreteRG}. The relation is exact only asymptotically, with corrections that decrease at deeper levels of the hierarchy.

The general solution of the discrete scaling relation can be written as
\begin{equation}
    \lambda_{\max}(d)=d^{\nu_{\rm HR}}
P\!\left(\frac{\ln d}{\ln\delta_F}\right),
\qquad
P(x+1)=P(x),\label{eq:periodicP}
\end{equation}
so that the Huberman--Rudnick power law and its log-periodic corrections are two consequences of the same discrete renormalization symmetry~\cite{Sornette1998}. Keeping the leading harmonic gives
\begin{equation}
\lmax(d)\simeq C\,d^{\nu_{\rm HR}}\left[1+b_1\cos\!\left(\frac{2\pi}{\ln\delta_F}\ln d+\varphi_1\right)\right],
\label{eq:logperiodic}
\end{equation}
with the Feigenbaum log-period
\begin{equation}
\Delta_{\log}=\ln\delta_F\simeq1.541.
\label{eq:Dlog}
\end{equation}

Figure~\ref{fig:4}(b) tests Eq.~\eqref{eq:logperiodic} on a dense scan of
1000 values of $\omega_d$. With the log-period fixed to the Feigenbaum value
$\ln\delta_F$, the fit gives
$\nu=0.445\pm0.003$ (statistical error) and $b_1\simeq0.07$.
The fitted exponent is stable at the level $\nu\simeq0.44$--$0.45$ under
reasonable variations of the fitting window and is consistent with the
Huberman--Rudnick prediction $\nu_{\rm HR}\simeq0.4498$. The leading harmonic captures the main oscillatory pattern of the Lyapunov exponent, but not the full non-sinusoidal fine structure.

A more direct, fit-independent test is provided by the band-merging points
$\omega_k$, at which the $3\cdot2^k$-band attractor merges into
$3\cdot2^{k-1}$ bands. We locate each $\omega_k$ independently of the
Lyapunov data, as the point where the coherent $1/(3\cdot2^k)$ component
loses its long-time phase coherence. The steps visible in
Fig.~\ref{fig:4}(b) coincide with the independently determined $\omega_k$. Their logarithmic spacing,
\begin{equation}
\ln\!\left[
\frac{\omega_c-\omega_k}
     {\omega_c-\omega_{k+1}}
\right]
=
1.559\pm0.008,
\end{equation}
is close to the Feigenbaum prediction
$\ln\delta_F=1.541$, and the merging sequence extrapolates to the same
accumulation point as the flip sequence. Moreover, at the merging points the
Lyapunov exponent obeys Eq.~\eqref{eq:discreteRG} directly: from the
outermost merger inward,
\begin{equation}
\frac{\lambda_{\max}(\omega_k)}
     {\lambda_{\max}(\omega_{k+1})}
=
1.95,\;2.00,\;2.00,\;1.99,
\end{equation}
in close agreement with the parameter-free prediction of a factor of two.
Thus the geometric merger hierarchy and the Lyapunov scaling provide
independent evidence for Feigenbaum discrete scale invariance, while the
log-periodic modulation resolves the corresponding fine structure in the
dense scan.

Crucially, the onset of chaos does not coincide with the loss of the original period-three organization. At the accumulation point, the discrete periodic orbit gives way to a banded chaotic attractor whose bands remain cyclically permuted by the drive. Period doubling therefore provides a direct route from a discrete time crystal to a time semicrystal: strict periodic motion is lost at the onset of chaos, while a finite temporal skeleton survives within the chaotic attractor.

% =====================================================================
\section{Reorganization of banded chaos}
\label{sec:reorg}
% =====================================================================

\begin{figure}[t]
\centering
\includegraphics[width=\linewidth]{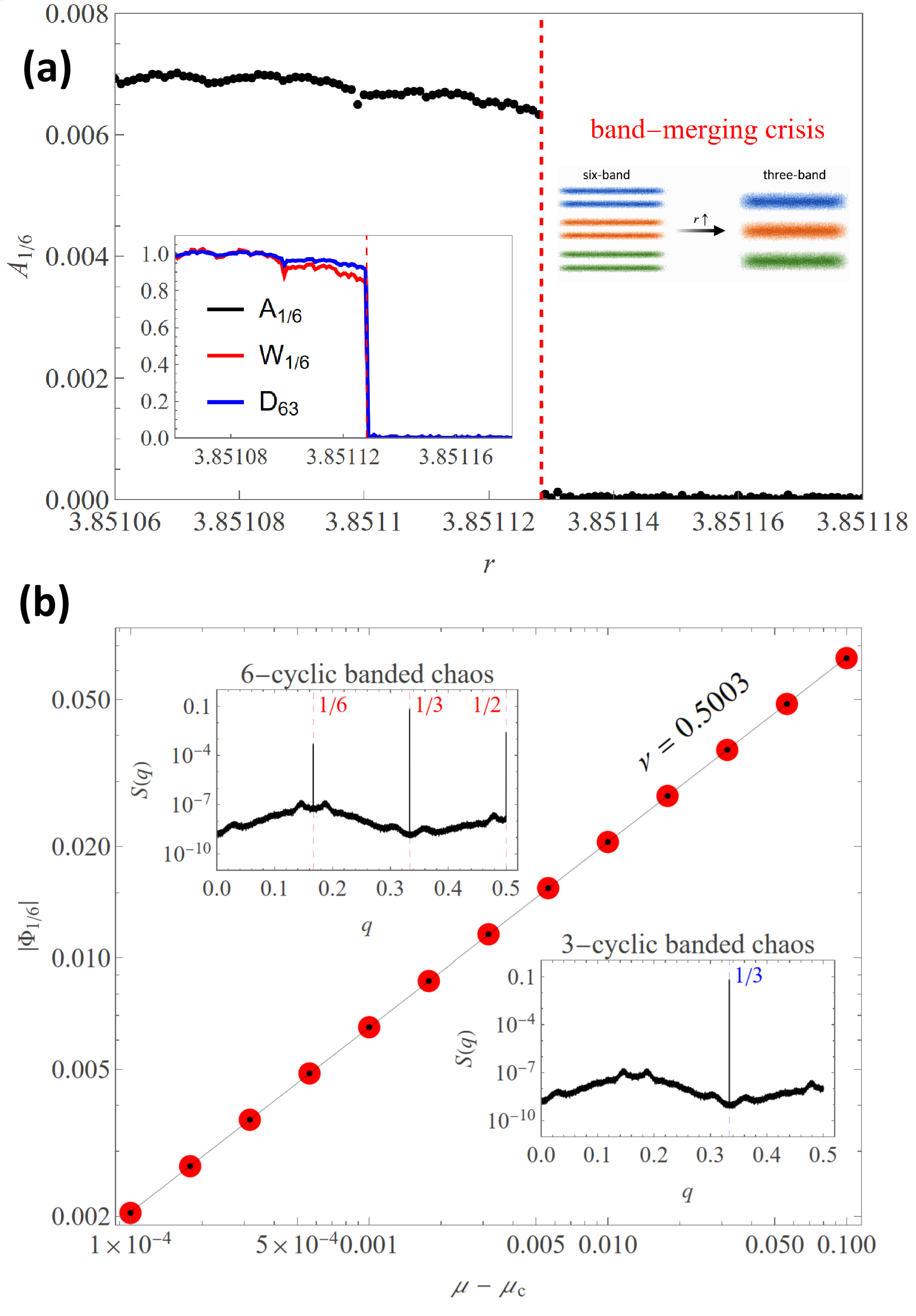}
\caption{
\textbf{Abrupt and continuous reorganization of banded chaos.}
\textbf{(a)} Abrupt $6\to3$ band-merging crisis in the logistic map.
The coherent $1/6$ amplitude $A_{1/6}=|\Phi_{1/6}|$ remains finite,
of order $7\times10^{-3}$, up to
$r_c=3.8511284$ and collapses at the crisis, while the underlying
$1/3$ temporal organization survives. The inset compares the normalized
$A_{1/6}$, spectral weight $W_{1/6}$, and geometric band-pair separation
$D_{63}$, which consistently identify the same loss of six-cyclic
organization.
\textbf{(b)} Continuous $3\to6$ reorganization in the coupled map,
Eq.~\eqref{eq:coupledmap}, for $r=3.853$, $c=0.5$, and $g=1$.
The emergent $1/6$ component follows
$|\Phi_{1/6}|\propto(\mu-\mu_c)^\nu$ over the fitting window
$10^{-4}\leq\mu-\mu_c\leq10^{-1}$, with
$\nu=0.5003\pm0.0003$, consistent with the square-root value
$\nu=1/2$. The insets show the stroboscopic spectra $S(q)$ below and
above the instability. Three-cyclic banded chaos contains the coherent
$q=1/3$ line, whereas above threshold a new $q=1/6$ component appears
together with a strong $q=1/2$ line, while the broadband chaotic
background remains present.
}
\label{fig:5}
\end{figure}

Banded-chaotic states can reorganize their temporal structure without losing chaos. In particular, the number of cyclically permuted bands can change while the largest Lyapunov exponent remains positive on both sides of the transition. Such transitions within the time-semicrystalline regime need not proceed through a unique mechanism. Figure~\ref{fig:5} illustrates two contrasting cases: an abrupt band-merging crisis and a continuous band-doubling transverse instability.

We first consider the logistic map inside the banded-chaotic regime. Near $r_c\simeq3.8511284$, the six-band chaotic attractor undergoes a
pairwise merger of its bands, reducing the cyclic structure from six bands
to three~\cite{Brown1986,Romeiras1988}. For the logistic map the state is simply the scalar variable $z_n=x_n$.
In the six-band regime, successive iterations visit the six bands in a fixed
cyclic order. We therefore group the trajectory according to $n \bmod 6$
and define $\bar z_j=\langle z_{6\ell+j}\rangle_\ell$ as the average position
of the $j$th band.

The $6\to3$ transition occurs when bands separated by three steps in this
cycle become indistinguishable. We quantify this pairwise separation through
\begin{equation}
D_{63}^{\rm raw}\equiv\frac{1}{3}\sum_{j=0}^{2}
\left\|\bar z_{j+3}-\bar z_j\right\| .
\label{eq:D63}
\end{equation}
Thus $D_{63}^{\rm raw}$ measures the average distance between the three band
pairs that merge at the transition, or in other words the distance between their centers of mass. It is finite in the six-band state and
vanishes when each pair has merged, leaving only three distinct bands.
For ease of comparison with the spectral quantities, we plot the normalized
value $D_{63}=D_{63}^{\rm raw}/D_{63}^{\rm ref}$, where
$D_{63}^{\rm ref}$ is a reference value taken in the six-band regime.

As shown in Fig.~\ref{fig:5}(a), the coherent $1/6$ amplitude remains finite up to the crisis and then collapses abruptly. The inset shows the same transition in the normalized coherent amplitude $A_{1/6}$, its spectral weight $W_{1/6}$, and the geometric separation $D_{63}$. The $1/3$ component, by contrast, is essentially unchanged across the merger, $A_{1/3}\simeq0.236$ on both sides of $r_c$. The crisis therefore removes the finer sixfold temporal organization while preserving the underlying threefold skeleton,
\[
\text{6-cyclic chaos}\longrightarrow\text{3-cyclic chaos},
\]
without destroying chaos itself.

This abrupt scenario is not the only possible route between time-semicrystalline states. To construct a continuous counterpart, we couple a three-band chaotic map to a transverse period-doubling degree of freedom,
\begin{equation}
\begin{aligned}
x_{n+1} &= r x_n(1-x_n),\\
y_{n+1} &= -\left[1+\mu+c\left(x_n-\tfrac{1}{2}\right)\right]y_n+g\,y_n^3 .
\end{aligned}
\label{eq:coupledmap}
\end{equation}
The logistic sector is kept in the three-band chaotic regime, while the transverse variable $y_n$ is multiplicatively modulated by the chaotic trajectory $x_n$. The state $y_n=0$ is an exact solution of Eq.~\eqref{eq:coupledmap}. In this
state the dynamics is entirely controlled by the three-band chaotic logistic
map $x_n$. To determine when a nonzero $y_n$ response can develop, we consider
a small deviation $\delta y_n$ from $y=0$. To linear order,
\begin{equation}
\delta y_{n+1}
=
-\left[1+\mu+c\left(x_n-\tfrac{1}{2}\right)\right]\delta y_n .
\end{equation}
The factor in brackets therefore determines whether a small deviation from
$y=0$ grows or shrinks at each iteration. Since $x_n$ is chaotic, this factor
fluctuates in time, and its average growth rate is measured by
\begin{equation}
\lambda_\perp(\mu)=
\left\langle
\ln\left|1+\mu+c\left(x_n-\tfrac{1}{2}\right)\right|
\right\rangle .
\label{eq:lambdaperp}
\end{equation}
When $\lambda_\perp<0$, small deviations decay and the system returns to
$y=0$. When $\lambda_\perp>0$, they instead grow and a finite $y_n$ response
develops. The transition therefore occurs when $\lambda_\perp$ crosses zero. For $r=3.853$ and $c=0.5$, this occurs at $\mu_c\simeq-1.85\times10^{-3}$. In Eq.~\eqref{eq:lambdaperp}, $\langle\cdots\rangle$ denotes the average over the natural measure of the three-band logistic attractor, equivalently the long-time average $\lim_{N\to\infty}N^{-1}\sum_{n=0}^{N-1}(\cdots)$ along a trajectory $x_n$.

Above threshold, a transverse period-two component develops and doubles the cyclic temporal structure of the three-band chaotic state. For the observable
\begin{equation}
O_n=x_n(1+y_n),
\label{eq:Ocoupled}
\end{equation}
this produces a coherent $1/6$ component on top of the broadband chaotic background. Over the resolved scaling regime, Fig.~\ref{fig:5}(b) gives
$|\Phi_{1/6}|\propto(\mu-\mu_c)^{0.5003}$, in excellent agreement with the
expected square-root onset. Very close to the transition, chaotic fluctuations lead to intermittent
behavior and deviations from the simple square-root scaling; see the
Supplemental Material~\cite{PlattSpiegelTresser1993,OttSommerer1994}.

The spectral insets make the reorganization explicit. Below the instability, the chaotic spectrum contains the coherent $1/3$ line characteristic of three-cyclic banded chaos. Above threshold, a new $1/6$ component appears while the broadband chaotic background remains present. A strong $1/2$ line is also generated by the product structure of the observable: the nonzero mean component of $x_n$ couples directly to the period-two oscillation of $y_n$. The transition therefore reorganizes
\[
\text{3-cyclic chaos}\longrightarrow\text{6-cyclic chaos},
\]
without destroying chaos itself.

The two examples demonstrate that transitions between
time-semicrystalline states need not proceed through a unique dynamical
mechanism. The cyclic organization can change discontinuously through a global band-merging crisis, or continuously through a transverse instability of a chaotic invariant set. In the latter case, square-root scaling is recovered away from the immediate threshold, while multiplicative chaotic fluctuations generate on--off-intermittency corrections asymptotically close to it. What is common to both cases is that the transition reorganizes the temporal skeleton of an already chaotic state rather than creating or destroying chaos itself.

These examples also show that there is no universal scaling law for the
reorganization of temporal order within chaos. Depending on the dynamical
mechanism, the change can be abrupt or continuous, and in the latter case
the critical scaling is controlled by the nonlinear structure of the
instability. The square-root behavior found above is therefore a property
of the present model rather than a universal exponent. More general
transverse maps, with competing nonlinear terms, can display different
scaling and crossover exponents, including effective values close to that
reported for the holographic time semicrystal. This observation will be
useful when comparing with the holographic result in Sec.~\ref{sec:holo}.

% =====================================================================
\section{From banded chaos to single-band chaos}
\label{sec:loss}
% =====================================================================

We finally consider how the residual temporal order of a time semicrystal is
destroyed. In the Duffing oscillator this occurs through an attractor-merging
crisis, providing a direct realization of the final
time-semicrystal-to-chaos transition in a continuously driven system.
Importantly, the crisis does not mark the onset of chaos: the dynamics is
already chaotic on both sides. What changes is the finite temporal structure
carried by the chaotic attractor.

For the Duffing parameters used in Fig.~\ref{fig:2},
$\alpha=-1$, $\beta=1$, $\gamma=0.30$, and $F=0.58$, the realization of the
spatiotemporal $\mathbb{Z}_2$ symmetry changes more than once along the
banded-chaotic branch. Immediately after the period-doubling cascade, the
chaotic attractor occurs as a symmetry-related mirror pair. Near
$\omega_d\simeq1.35035$, these two attractors merge into a single
$\mathcal S$-symmetric three-band chaotic attractor. This is the regime
containing the representative state shown in Fig.~\ref{fig:2} at
$\omega_d=1.36$. Upon increasing the driving frequency further, the
symmetric banded attractor becomes a mirror-related pair again near
$\omega_d\simeq1.3874$. Across both events the maximal Lyapunov exponent
remains positive and the coherent $1/3$ component remains essentially
unchanged: numerically,
$|\Phi^{\rm ct}_{1/3}|\simeq0.63$ across the first event and
$|\Phi^{\rm ct}_{1/3}|\simeq0.62$ across the second. Only the
symmetry-breaking order parameter changes between zero and a finite value.
These are therefore band-preserving symmetry-changing events rather than
transitions that destroy the temporal skeleton
\cite{ChossatGolubitsky1988}.

To quantify separately the temporal and symmetry structures, we use the
continuous-time Fourier amplitude
\begin{equation}
\Phi^{\rm ct}_{1/3}
=
\lim_{T\to\infty}
\frac{1}{T}
\int_0^T
x(t)\,e^{-i\omega_d t/3}\,dt .
\label{eq:Phict}
\end{equation}
Its magnitude is identical for the two mirror-related attractors and
therefore provides a symmetry-independent measure of the period-three
temporal order. The time-averaged displacement
$\bar{x}\equiv\langle x(t)\rangle_t$, by contrast, changes sign between the
two members of a mirror pair and vanishes for an
$\mathcal S$-symmetric attractor. It therefore acts as an order parameter for
the broken spatiotemporal $\mathbb{Z}_2$ symmetry. The two quantities probe
different aspects of the dynamics: $\bar{x}$ diagnoses the realization of
the symmetry, whereas $|\Phi^{\rm ct}_{1/3}|$ diagnoses the persistence of
the threefold temporal skeleton.

\begin{figure}[t]
\centering
\includegraphics[width=\columnwidth]{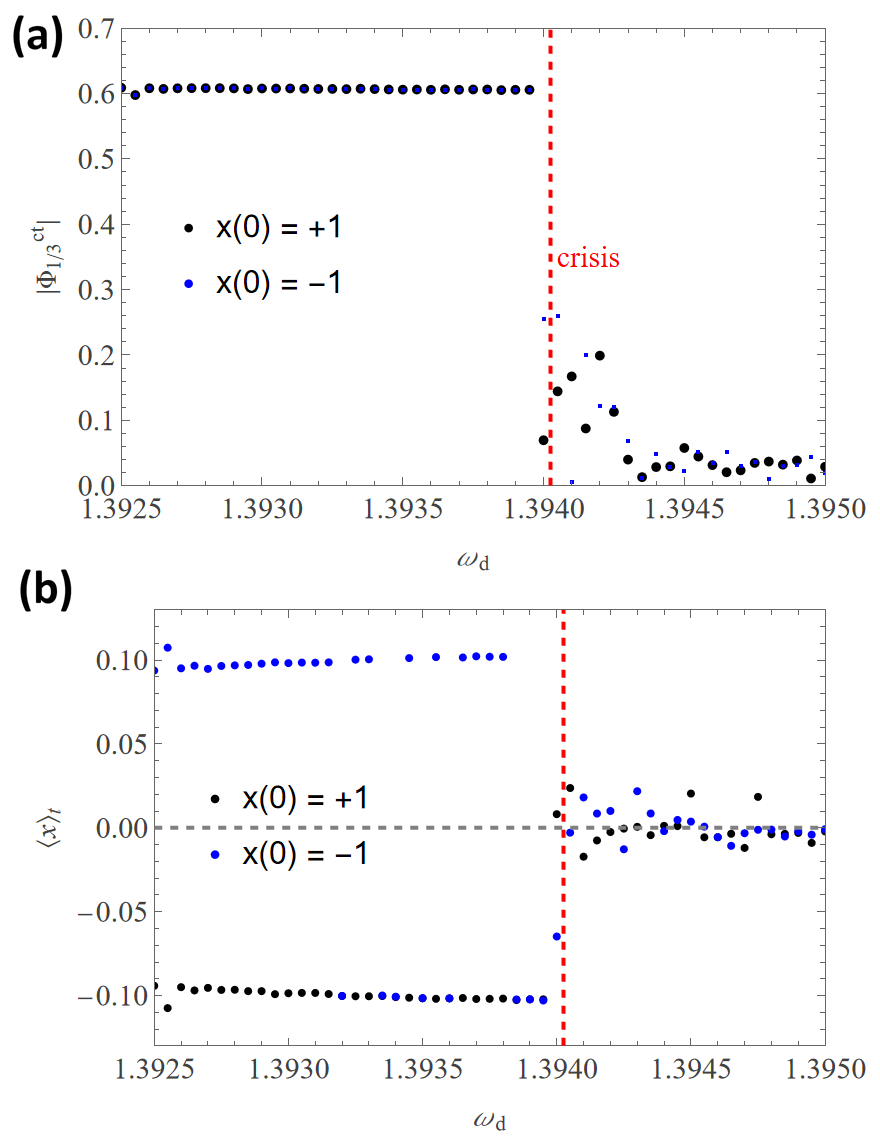}
\caption{
\textbf{Loss of time-semicrystalline order through an attractor-merging
crisis in the driven Duffing oscillator.}
The driving frequency $\omega_d$ is varied at fixed
$\alpha=-1$, $\beta=1$, $\gamma=0.30$, and $F=0.58$ across the final crisis
at $\omega^\star\simeq1.39399$ (red dashed line).
\textbf{(a)} Coherent continuous-time subharmonic amplitude
$|\Phi^{\rm ct}_{1/3}|$. On the time-semicrystalline side the $1/3$
component remains finite, $|\Phi^{\rm ct}_{1/3}|\simeq0.60$, and collapses
at the crisis.
\textbf{(b)} Symmetry-breaking order parameter $\bar{x}$ for initial
conditions $(x_0,\dot{x}_0)=(+1,0)$ and $(-1,0)$. Just below the crisis the
two mirror-related chaotic attractors carry opposite nonzero values
$\bar{x}\simeq\pm0.10$; at the crisis they merge and $\bar{x}$ collapses to
zero.
}
\label{fig:6}
\end{figure}

Immediately below the final crisis,
$|\Phi^{\rm ct}_{1/3}|\simeq0.60$ and
$\bar{x}\simeq\pm0.10$. As the driving frequency reaches
\begin{equation}
1.39398<\omega^\star<1.39399,
\end{equation}
the two chaotic attractors collide and merge. Figure~\ref{fig:6} shows that
this event is qualitatively different from the preceding symmetry changes:
the merger not only restores the spatiotemporal symmetry,
$\bar{x}\rightarrow0$, but also joins the remaining three chaotic bands and
destroys the coherent $1/3$ component,
$|\Phi^{\rm ct}_{1/3}|\rightarrow0$. Symmetry restoration and loss of
temporal order are therefore distinct phenomena. They occur separately at
the lower-frequency band-preserving events and coincide only at the final
band-destroying crisis.

Close to the final merger, the two chaotic regions are already very close in
phase space, and trajectories can spend long times near one of them before
moving to the other. As a result, finite-time measurements may still show a
small remnant of the former subharmonic response even after the transition.
In the spectrum this appears as a narrow peak rather than a truly persistent
sharp line. Therefore, a residual peak observed over a finite time does not
necessarily imply that the corresponding temporal order survives at long
times~\cite{Brown1986,Romeiras1988,GrebogiOttRomeirasYorke1987}. Because the two chaotic states have nearby and intertwined regions of
attraction, the same initial condition does not always select the same branch
as $\omega_d$ is varied, which explains the occasional branch switching in
Fig.~\ref{fig:6}(b).

The distinction between symmetry restoration and loss of temporal order also
has a direct spectral interpretation. An
$\mathcal S$-symmetric three-band chaotic attractor can still retain the
coherent subharmonic lines associated with the threefold band permutation.
The symmetry selection rule removes only the components forbidden by
$\mathcal S$; it does not by itself eliminate the $1/3$ temporal skeleton.
After the final crisis, by contrast, the three-band organization itself is
lost and the coherent $1/3$ component disappears. The spectral structure is
therefore a more direct diagnostic of the survival of time-semicrystalline
order than the symmetry-breaking parameter $\bar{x}$ alone.

\begin{figure*}[t]
\centering
\includegraphics[width=\textwidth]{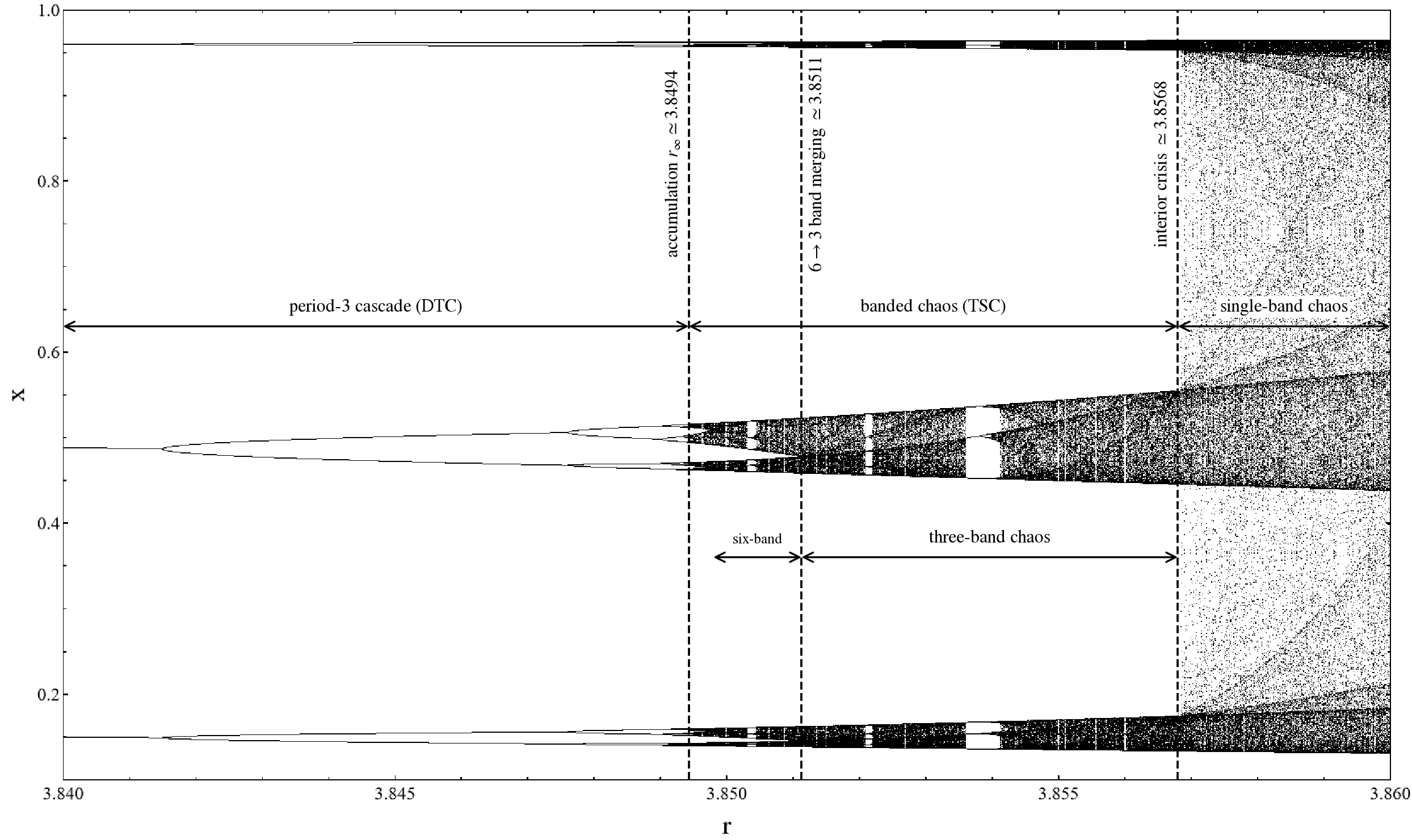}
\caption{
\textbf{From periodic order to single-band chaos in the logistic map.}
Starting from the stable period-$3$ orbit, the system undergoes the
period-doubling cascade $3\to6\to12\to\cdots$, accumulating at
$r_\infty\simeq3.8494$. Beyond the accumulation point the dynamics is
chaotic but remains organized into multiple cyclically visited bands,
corresponding to the time-semicrystalline regime. Successive band-merging
crises reduce the number of chaotic bands; representative
$12\to6$ and $6\to3$ mergers are indicated. The three-band chaotic attractor
persists until the interior crisis at $r\simeq3.8568007$, where it expands
abruptly into a single-band chaotic attractor and the remaining cyclic
temporal organization is lost.
}
\label{fig:7}
\end{figure*}

The same separation between the onset of chaos and the loss of temporal
organization can be resolved particularly transparently in the logistic
map. Its period-three window contains the complete sequence from periodic
subharmonic order, through the Feigenbaum cascade and banded chaos, to the
final destruction of cyclic organization. Figure~\ref{fig:7} summarizes
this evolution.

Starting from the stable period-$3$ orbit, increasing $r$ generates the
period-doubling cascade
\begin{equation}
3\to6\to12\to\cdots,
\label{eq:cascadeLogistic}
\end{equation}
which accumulates at $r=r_\infty\simeq3.8494$. At the accumulation point the
limiting attractor is the critical Feigenbaum set and the Lyapunov exponent
vanishes. Immediately beyond it, the dynamics becomes chaotic, but the
attractor remains divided into several cyclically visited bands. Chaos has
therefore developed without destroying the finite temporal skeleton: this is
the banded-chaotic, or time-semicrystalline, regime.

As $r$ is increased further, the chaotic bands successively merge. These
events progressively remove the finer levels of cyclic temporal organization
while leaving coarser ones intact. For example, the six-band attractor
undergoes a $6\to3$ band-merging crisis at
$r\simeq3.8511284$, after which the threefold cyclic structure survives even
though the sixfold component has disappeared. This provides a second example
of a transition that reorganizes an already chaotic attractor without
destroying all of its temporal order.

The subsequent transition near $r\simeq3.8568007$ is qualitatively different.
Here the three-band chaotic attractor suddenly expands and becomes a single
broad chaotic attractor after colliding with a nearby unstable orbit. This
abrupt event, known as an \emph{interior crisis}, destroys the remaining
threefold cyclic organization altogether. Unlike the preceding $6\to3$
merger, no coarser temporal skeleton survives.

In spectral terms, the successive band mergings remove progressively finer
coherent subharmonic components while leaving coarser components intact,
whereas the final interior crisis destroys the last residual pure-point
contribution associated with the three-cyclic structure. The transition to
single-band chaos is therefore not the onset of chaos itself, nor is it
identified simply by restoration of a discrete symmetry. It is the loss of
the remaining temporal skeleton within an already chaotic state.
% =====================================================================
\section{Connection to the holographic time semicrystal}
\label{sec:holo}
% =====================================================================

The banded-chaos picture provides a direct point of comparison with the holographic time semicrystal of Ref.~\cite{Lei2026}. At the onset of chaos, that work reports $\lmax\sim d^{\nu}$ with $\nu=0.450\pm0.009$, remarkably close to the Huberman--Rudnick value $\nu_{\rm HR}\simeq0.4498$ of Eq.~\eqref{ruru}. This quantitative agreement suggests that the first transition---from periodic subharmonic order to a chaotic state retaining a finite temporal skeleton---is governed by the same period-doubling mechanism. A sharper test follows from Fig.~\ref{fig:4}(b): if the holographic onset is governed by the Feigenbaum hierarchy, the Lyapunov exponent should carry a log-periodic modulation of log-period $\ln\delta_F$, i.e., of frequency $\Omega_F=2\pi/\ln\delta_F\simeq4.08$ in the convention of Ref.~\cite{Lei2026}, with steps located where the successively finer subharmonic components lose their coherence.

The subsequent reorganization of the temporal skeleton appears to involve
different physics. Reference~\cite{Lei2026} finds that, close to the
transition at $\omega_s$, the amplitude $A$ characterizing the surviving
subharmonic temporal order scales as
\begin{equation}
A\propto(\omega_s-\omega_d)^\nu,
\qquad
\nu=0.360\pm0.017 ,
\label{eq:holoexp}
\end{equation}
together with log-periodic oscillations of frequency
$\Omega=16.4\pm0.7$. The latter correspond to a discrete scale factor
\begin{equation}
\Lambda=\exp\!\left(\frac{2\pi}{\Omega}\right)\simeq1.47 .
\label{eq:Lambda}
\end{equation}
This scale is very different from the Feigenbaum factor
$\delta_F\simeq4.67$, or equivalently from the Feigenbaum log-periodic
frequency $\Omega_F=2\pi/\ln\delta_F\simeq4.08$ found at the onset of chaos.
Thus, although the holographic oscillations may still signal a form of
discrete scale invariance, they do not appear to originate from the
Feigenbaum period-doubling hierarchy discussed above.

There is no inconsistency between the reported exponent $\nu\simeq0.36$ and
the oscillation period: in a generic discrete-scale-invariant system these
quantities need not be related unless an additional dynamical mechanism
imposes such a relation. Our results likewise show that reorganizations of
temporal order within chaos need not have a universal scaling exponent.
The holographic transition may therefore represent a different dynamical
mechanism, whose underlying hierarchy remains to be identified.

The comparison is particularly direct because the holographic model possesses
the same spatiotemporal $\mathbb{Z}_2$ structure. Its scalar potential is even
under $\phi\to-\phi$, while the periodic drive
$F_d(t)=F_0\sin(\omega_d t)$ enters linearly in the boundary condition of
the $\mathbb{Z}_2$-odd scalar field. Since
$F_d(t+T_d/2)=-F_d(t)$, the driven equations and boundary condition are
invariant under
\begin{equation}
\mathcal{S}:\qquad
X(t)\longrightarrow -X(t+T_d/2),
\end{equation}
with $X(t)$ the boundary condensate. The spectral selection rule derived
above for the Duffing oscillator therefore applies directly to the
holographic response.

The same symmetry provides a direct spectral diagnostic. For a
$\mathcal{S}$-symmetric three-cyclic state, the coherent response satisfies
$X(t+3T_d/2)=-X(t)$, so only odd multiples of $\omega_d/3$ are allowed in
the coherent spectrum. In particular, the DC component and the
$2\omega_d/3$ harmonic must vanish, while the $\omega_d/3$ component may
remain finite. This distinguishes restoration of the spatiotemporal
$\mathbb{Z}_2$ symmetry from destruction of the threefold temporal skeleton
without requiring reconstruction of the underlying attractor.

% =====================================================================
\section{Discussion and outlook}
\label{sec:discussion}
% =====================================================================

Time-semicrystalline order does not require a new dynamical mechanism. A chaotic attractor whose bands are cyclically permuted by the drive already supports exact subharmonic order, even though the motion within each band is unpredictable. From this viewpoint, the onset of chaos and the loss of temporal order are distinct events. Chaos emerges through the period-doubling cascade, whereas temporal order is ultimately destroyed by band mergers and crises. The time-semicrystalline regime is therefore identified by the coexistence of a positive Lyapunov exponent and a coherent subharmonic component that remains finite at long times. Neither property alone is sufficient.

Although we have focused on the Duffing oscillator and the logistic map as controlled benchmark systems, the underlying mechanism is not specific to either model. Any dissipative driven system in which a subharmonic orbit enters chaos through period doubling may, in principle, retain its original cyclic organization as a banded chaotic attractor. We therefore expect analogous phenomenology in other nonlinear oscillators, such as driven Van der Pol systems, and potentially in spatially extended driven models such as Ginzburg--Landau fields, where the cyclic band structure could acquire the character of a collective temporal order parameter.

This picture also separates what is fragile from what is robust. Periodic windows are interspersed throughout the banded-chaotic regime, so chaos itself can be lost under small parameter variations. The underlying temporal organization is more persistent: periodic windows within the same interval retain the corresponding subharmonic structure, while chaotic regions preserve it through the cyclic permutation of the attractor bands. The robust feature of a time semicrystal is therefore the temporal skeleton, rather than chaos itself. More generally, systems belonging to the same period-doubling universality class should exhibit analogous behavior, including Feigenbaum scaling of the Lyapunov exponent and the associated log-periodic corrections.

An important extension concerns the thermal melting of time-semicrystalline order. The present analysis is deterministic and therefore does not address how the cyclic temporal skeleton survives in the presence of fluctuations. A simple first test within the Duffing setting is to increase the damping strength and determine how the banded-chaotic regime and its bounding crises evolve under stronger dissipation, although damping alone should not be identified with a finite temperature. A genuine thermal extension would require stochastic forcing, ideally constrained by fluctuation--dissipation. In a single oscillator, noise gives subharmonic states a finite lifetime and is known to erase the finer levels of the band hierarchy first~\cite{Crutchfield1982}; it would therefore be interesting to determine how the lifetime of the surviving temporal order scales with noise strength and with the distance from the bounding crises. Holographic models provide a particularly natural complementary setting, since temperature is an intrinsic thermodynamic control parameter while strongly nonlinear driven dynamics and chaos remain directly accessible. They may therefore offer a controlled route to study the finite-temperature melting of time-semicrystalline order and to determine whether the temporal skeleton is lost through broadened band-merging phenomena, activated switching between temporal sectors, or qualitatively new mechanisms.

The classical interpretation developed here also makes time-semicrystalline
phenomenology accessible to relatively simple laboratory experiments.
Mechanical and electronic realizations of Duffing-type dynamics have long
provided tabletop platforms in which subharmonic responses, period-doubling
cascades, and deterministic chaos can be observed directly
~\cite{Novak1982,Berger1997,Jones2001}. In particular, electronic Duffing
analogues have reproduced the Feigenbaum route to chaos, while similar
period-doubling scaling has also been measured in nonlinear electronic
circuits~\cite{Jones2001,Hanias2009}. Such systems therefore provide a
natural setting in which to test the central prediction of the present work:
that the onset of chaos need not destroy subharmonic temporal order, but can
instead produce a banded-chaotic regime in which coherent subharmonic
components coexist with a positive Lyapunov exponent. Simultaneous
measurements of the stroboscopic attractor, subharmonic spectral weight, and
Lyapunov exponent across the cascade could provide a direct tabletop
realization of the transition from periodic time-crystalline dynamics to
time-semicrystalline order and, eventually, to single-band chaos.

A final open question concerns many-body systems. Coupling many such units could promote the band index to a collective order parameter. Classical extended systems can sustain macroscopic temporal order despite local chaos or noise~\cite{Yao2020,Bohr1987,Bennett1990,Chate1992}, suggesting that arrays of banded-chaotic units may support a genuine time-semicrystalline phase, together with a collective transition associated with the loss of its temporal order~\cite{Yue2022}.

A related direction concerns the recently proposed time-glass phase, in which spatial symmetry breaking coexists with chaotic rather than periodic dynamics of a collective order parameter~\cite{Haga2026}. The present framework suggests a simple classical route toward analogous behavior. Arrays of coupled driven nonlinear oscillators could develop collective symmetry breaking while their synchronized macroscopic motion becomes chaotic. Such a state would differ sharply from the time semicrystal discussed here: banded chaos retains a finite cyclic temporal skeleton and hence coherent subharmonic components, whereas a time-glass analogue would display collective chaos without persistent subharmonic order. Coupled Duffing or Van der Pol oscillators therefore provide a natural minimal setting in which to explore the relation and possible transitions between time-crystalline, time-semicrystalline, and time-glass dynamics.

\begin{acknowledgments}
MB thanks Y.Q.~Lei, X.H.~Ge and Y.~Tian for explaining their results in Ref.~\cite{Lei2026}. MB thanks Z.~Cai, P.~Yang and in particular X.H.~Ge for useful discussions and comments on a first version of this manuscript. MB acknowledges the support of the Foreign Young Scholars Research Fund Project (Grant No.~22Z033100604) and the sponsorship from the Yangyang Development Fund. The author used ChatGPT (OpenAI) and Claude Opus 5.5 (Anthropic) as AI-assisted tools for language editing, assistance with coding and debugging, and independent numerical cross-checks of selected results. The schematic in Fig.~\ref{fig:1}(a) was designed by the author and rendered with assistance from ChatGPT. The scientific question, conceptual framework, and interpretation of the results were formulated by the author, who checked all AI-assisted output and takes full responsibility for the content of this manuscript.
\end{acknowledgments}

\clearpage
\onecolumngrid

\section*{Supplemental Material}

% =====================================================================
% Supplemental Material numbering
% =====================================================================

\setcounter{equation}{0}
\setcounter{figure}{0}
\setcounter{table}{0}
\setcounter{section}{0}

\renewcommand{\theequation}{S\arabic{equation}}
\renewcommand{\thefigure}{S\arabic{figure}}
\renewcommand{\thetable}{S\arabic{table}}
\renewcommand{\thesection}{S\arabic{section}}

\section{Numerical methods and diagnostics}
\label{sec:SMnumerics}

This section summarizes the numerical procedures used throughout the main
text. Model parameters and control-parameter ranges specific to individual
calculations are given in the corresponding figure captions. 

\subsection{Duffing dynamics and stroboscopic sampling}

The driven Duffing oscillator is written as the first-order system
\begin{equation}
\dot{x}=v,
\qquad
\dot{v}
=
-\gamma v-\alpha x-\beta x^3+F\cos(\omega_d t).
\label{eq:SMDuffingFirstOrder}
\end{equation}
Trajectories are obtained by direct numerical integration of
Eq.~\eqref{eq:SMDuffingFirstOrder} with a fixed-step fourth-order
Runge--Kutta scheme. We use $n_s=200$ integration steps per driving period
for the calculations associated with Figs.~\ref{fig:1}, \ref{fig:2}, and \ref{fig:6} of the main text,
while the higher-resolution Feigenbaum analysis of Fig.~\ref{fig:4} uses
$n_s=400$ steps per period.

For each value of the control parameter, an initial transient is discarded
before any observable is evaluated. The stroboscopic trajectory is then
constructed by sampling once per driving period,
\begin{equation}
{\bf z}_n
=
\left[
x(nT_d),v(nT_d)
\right],
\qquad
T_d=\frac{2\pi}{\omega_d}.
\label{eq:SMstrobe}
\end{equation}
Periodic cycles, chaotic bands, and their cyclic permutation are identified
from these post-transient stroboscopic states. Unless stated otherwise,
initial conditions are those specified in the corresponding figure caption,
since different attractors may coexist in parts of parameter space.

Whenever coherent continuous-time Fourier components are required, we
evaluate them directly on the post-transient trajectory rather than inferring
them from the height of a discrete Fourier-transform bin. In particular,
\begin{equation}
\Phi_q^{\rm ct}
=
\frac{1}{T}
\int_{t_0}^{t_0+T}
x(t)\,
e^{-iq\omega_d t}\,dt ,
\label{eq:SMcontinuousPhi}
\end{equation}
with $t_0$ chosen after the transient. For the period-three component used in
Fig.~\ref{fig:6}, $q=1/3$. The observation interval $T$ is chosen as an integer
multiple of $3T_d$ and, more generally, of $pT_d$ when measuring a component
$q=1/p$, preventing leakage between the drive harmonics and the target
subharmonic component.

\subsection{Largest Lyapunov exponent}

The largest Lyapunov exponent of the Duffing dynamics is evaluated by
integrating the tangent-space equations together with the reference
trajectory. Linearizing Eq.~\eqref{eq:SMDuffingFirstOrder} gives
\begin{equation}
\frac{d}{dt}
\begin{pmatrix}
\delta x\\
\delta v
\end{pmatrix}
=
\begin{pmatrix}
0 & 1\\
-\alpha-3\beta x(t)^2 & -\gamma
\end{pmatrix}
\begin{pmatrix}
\delta x\\
\delta v
\end{pmatrix}.
\label{eq:SMtangent}
\end{equation}
We use the standard Benettin procedure~\cite{Benettin1980}. After each
renormalization interval $\tau_R=T_d$, the tangent vector is rescaled to unit
norm and the logarithmic stretching is recorded. After $M$
renormalizations,
\begin{equation}
\lambda_{\max}
=
\frac{1}{M\tau_R}
\sum_{k=1}^{M}
\ln
\frac{
\|\delta{\bf z}(k\tau_R)^-\|
}{
\|\delta{\bf z}((k-1)\tau_R)^+\|
},
\label{eq:SMBenettin}
\end{equation}
where the superscripts $-$ and $+$ denote the tangent vector immediately
before and after renormalization, respectively. Thus $\lambda_{\max}$ is
reported per unit physical time. Statistical uncertainties are estimated by
dividing the accumulated stretching factors into consecutive blocks and
taking the standard error of the corresponding block averages.

\subsection{Logistic-map calculations}

For the logistic map,
\begin{equation}
x_{n+1}=r x_n(1-x_n),
\label{eq:SMlogistic}
\end{equation}
no time integration is required: after discarding an initial transient, the
trajectory is obtained simply by repeated application of the map.

The Lyapunov exponent measures how a small separation between two nearby
trajectories grows or shrinks under these iterations. Let $\delta x_n$ be a
small perturbation of $x_n$. Linearizing Eq.~\eqref{eq:SMlogistic} gives
\begin{equation}
\delta x_{n+1}
=
f'(x_n)\,\delta x_n
=
r(1-2x_n)\,\delta x_n ,
\end{equation}
where $f(x)=rx(1-x)$. After $N$ iterations, the perturbation is therefore
\begin{equation}
|\delta x_N|
=
|\delta x_0|
\prod_{n=0}^{N-1}
\left|r(1-2x_n)\right|.
\end{equation}
Taking the logarithm converts this product into a sum,
\begin{equation}
\frac{1}{N}
\ln\frac{|\delta x_N|}{|\delta x_0|}
=
\frac{1}{N}
\sum_{n=0}^{N-1}
\ln\left|r(1-2x_n)\right|.
\end{equation}
The long-time average growth rate of the perturbation is thus
\begin{equation}
\lambda_{\rm map}
=
\lim_{N\rightarrow\infty}
\frac{1}{N}
\sum_{n=0}^{N-1}
\ln\left|r(1-2x_n)\right|.
\label{eq:SMlogisticLE}
\end{equation}
This Lyapunov exponent is measured per iteration of the map. A positive
value means that initially nearby trajectories separate exponentially and
the dynamics is chaotic, whereas a negative value indicates convergence
toward a stable periodic orbit.
Periodic windows are identified either from the asymptotic orbit or from
$\lambda_{\rm map}<0$, whereas banded-chaotic regimes have
$\lambda_{\rm map}>0$ together with a finite long-time coherent component.
%The amplitudes of Eq.~\eqref{eq:SMdiscretePhi} and the block-coherence
%diagnostic of Eq.~\eqref{eq:SMcoherence} are applied directly to the sequence
%$x_n$.

Particularly long trajectories are required close to band-merging crises.
Finite-time values of subharmonic amplitudes and geometric band-pair
separations above a merger are therefore interpreted as residuals rather
than as asymptotic order. These residuals decrease systematically with
increasing observation time, consistently with the known spectral
broadening at chaotic band mergings
~\cite{Brown1986,Romeiras1988}.

\subsection{Numerical convergence}

Convergence was checked with respect to both temporal resolution and
trajectory length. Representative Duffing trajectories were repeated with
an increased number of integration steps per driving period; the resulting
attractor classification, Lyapunov exponents, coherent amplitudes, and
transition locations remained unchanged within the numerical accuracy
quoted in the text. The Feigenbaum analysis was performed at the higher
resolution $n_s=400$ because its innermost resolved scales lie only a few
parts in $10^{6}$ from the accumulation point.

The accumulation point entering Fig.~\ref{fig:4} is determined independently from the
resolved flip sequence and is
\begin{equation}
\omega_c
=
1.4349837\pm2\times10^{-7}.
\label{eq:SMomegac}
\end{equation}
The same accumulation point is obtained from the band-merging hierarchy
within this numerical uncertainty. The five successive merger distances and
their Lyapunov exponents are reported explicitly below, providing an
additional check that the innermost resolved scales are not numerical
artifacts.

Close to bifurcations and crises we increase the discarded transient and the
post-transient measurement interval until the relevant observables are stable
under a further increase of the trajectory length. Quoted errors on
Lyapunov exponents and coherent observables are statistical block errors,
while errors on fitted exponents are the corresponding fit uncertainties.
For the Huberman--Rudnick fit, variations of the fitting window give
$\nu\simeq0.44$--$0.45$, which is reported separately from the statistical
fit error. Transition points obtained by bisection are quoted only to the
precision for which the classification remains stable under these numerical
checks.
\section{Derivation of the line-plus-continuum spectrum}

Here we derive the spectral decomposition in Eq.~\eqref{eq:Sq} of the main text.
For a $p$-cyclic chaotic state, we write the stroboscopic observable as
\begin{equation}
O_n = a_{n\,({\rm mod}\,p)}+\delta O_n ,
\end{equation}
where $a_j$ is the conditional average of the observable in the $j$th
cyclic band, while $\delta O_n$ describes the fluctuations around this
periodic sequence.

Consider first the periodic contribution. For an observation time
$N=Mp$, its finite-time Fourier amplitude is
\begin{align}
\Phi_q^{(N)}
&=
\frac{1}{N}\sum_{n=0}^{N-1}
a_{n\,({\rm mod}\,p)} e^{-2\pi i q n}
\\
&=
\frac{1}{p}
\sum_{j=0}^{p-1}
a_j e^{-2\pi i q j}
\,
\frac{1}{M}
\sum_{\ell=0}^{M-1}
e^{-2\pi i q p\ell}.
\end{align}
In the long-time limit, the second factor vanishes unless
\begin{equation}
q=\frac{m}{p},
\qquad m=0,\ldots,p-1 ,
\end{equation}
for which it equals unity. The surviving Fourier amplitudes are therefore
\begin{equation}
\Phi_{m/p}
=
\frac{1}{p}
\sum_{j=0}^{p-1}
a_j e^{-2\pi i mj/p},
\end{equation}
as given in Eq.~\eqref{eq:Phimp} of the main text.

The exactly periodic motion between the $p$ cyclic bands therefore produces
sharp spectral lines at the reduced frequencies $q=m/p$, with weights
\begin{equation}
W_{m/p}=|\Phi_{m/p}|^2 .
\end{equation}

The remaining term $\delta O_n$ describes the irregular motion within the
bands. By construction it has zero conditional average within each cyclic
sector. If the dynamics of the $p$-fold map restricted to each band is
mixing, these fluctuations do not retain a persistent periodic component
and instead generate a continuous spectral background, which we denote by
$S_{\rm cont}(q)$.

The full spectral measure can therefore be written as
\begin{equation}
S(q)
=
\sum_{m=0}^{p-1}
W_{m/p}
\delta\!\left(q-\frac{m}{p}\right)
+
S_{\rm cont}(q),
\qquad q\in[0,1),
\end{equation}
which is Eq.~\eqref{eq:Sq} of the main text. Thus the sharp spectral lines originate
from the cyclic permutation of the bands, whereas the continuous background
comes from the chaotic dynamics within them. Additional discrete components
may appear if the intraband dynamics itself contains further coherent
frequencies.
\section{Duffing bifurcation structure and spatiotemporal symmetry}
Here we summarize the bifurcation structure of the Duffing branch used in
Figs.~\ref{fig:1}, \ref{fig:2}, and \ref{fig:6} of the main text and derive the spectral selection rule
associated with its spatiotemporal $\mathbb{Z}_2$ symmetry.

For an even restoring potential, the driven Duffing equation
\begin{equation}
\ddot{x}+\gamma\dot{x}+\alpha x+\beta x^3
=
F\cos(\omega_d t)
\end{equation}
is invariant under
\begin{equation}
\mathcal{S}:
\qquad
x(t)\longrightarrow
-x\!\left(t+\frac{T_d}{2}\right),
\qquad
T_d=\frac{2\pi}{\omega_d}.
\label{eq:SMSymmetry}
\end{equation}
Applying the transformation twice gives a translation by one driving
period, so that on stroboscopic states
\begin{equation}
\mathcal{S}^2=\mathcal{P},
\end{equation}
where $\mathcal{P}$ denotes the stroboscopic map. Periodic and chaotic
attractors may therefore either be invariant under $\mathcal{S}$ or occur
in symmetry-related pairs.

For the parameters
\begin{equation}
\alpha=-1,\qquad
\beta=1,\qquad
\gamma=0.30,\qquad
F=0.58,
\end{equation}
the period-three branch relevant to Fig.~\ref{fig:2} is created through a saddle-node
bifurcation at
\begin{equation}
\omega_d\simeq1.2821 .
\end{equation}
The stable branch is initially $\mathcal{S}$-symmetric. Upon increasing the driving frequency, the symmetric period-three solution
eventually becomes unstable at
\begin{equation}
\omega_d\simeq1.3218.
\end{equation}
Beyond this point, the original symmetric solution is replaced by two
equivalent period-three solutions related by the transformation
$\mathcal{S}$. In other words, the dynamics chooses one of two
symmetry-related branches and the $\mathcal{S}$ symmetry is spontaneously
broken. This splitting is the standard structure of a pitchfork
bifurcation. In the Floquet analysis, the same transition is signaled by a
real multiplier of the three-period return map $\mathcal{P}^3$ reaching
$+1$.

Each broken-symmetry branch subsequently undergoes the period-doubling
sequence
\begin{equation}
3\rightarrow6\rightarrow12\rightarrow24\rightarrow\cdots .
\end{equation}
The first few flip bifurcations occur at
\begin{equation}
\omega_d\simeq
1.33846,\quad
1.34330,\quad
1.34442,\quad
1.34467,
\end{equation}
where a real Floquet multiplier of the corresponding return map crosses
$-1$. Their accumulation near
\begin{equation}
\omega_d\simeq1.34473
\end{equation}
marks the onset of banded-chaotic dynamics on this branch. These values are
obtained by Newton continuation of the periodic solutions and direct
evaluation of the Floquet multipliers.

The symmetry has a simple consequence for the continuous-time spectrum.
For an $\mathcal{S}$-symmetric period-three state, advancing by half of its
full period, $3T_d/2$, reverses the sign of the response,
\begin{equation}
x\!\left(t+\frac{3T_d}{2}\right)=-x(t).
\label{eq:SMAntiperiodicity}
\end{equation}
Thus the second half of every $3T_d$ cycle is the sign-reversed copy of the
first half. The same relation applies to the coherent part of an
$\mathcal{S}$-symmetric three-band chaotic attractor. Expanding the coherent response
in harmonics of the fundamental subharmonic frequency,
\begin{equation}
x_{\rm coh}(t)
=
\sum_{m=-\infty}^{\infty}
c_m
e^{im\omega_d t/3},
\end{equation}
and substituting Eq.~\eqref{eq:SMAntiperiodicity} gives
\begin{equation}
c_m
e^{im\pi}
=
-c_m .
\end{equation}
Hence
\begin{equation}
c_m=0
\qquad
\text{for even }m,
\end{equation}
whereas odd harmonics are allowed. The coherent spectrum of the symmetric
period-three state can therefore contain
\begin{equation}
\frac{\omega_d}{3},
\quad
\omega_d,
\quad
\frac{5\omega_d}{3},
\quad\ldots,
\end{equation}
while the DC component and the even harmonics
\begin{equation}
\frac{2\omega_d}{3},
\quad
\frac{4\omega_d}{3},
\quad\ldots
\end{equation}
are forbidden. Once the $\mathbb{Z}_2$ symmetry is broken, this restriction
is lifted and even harmonics, together with a finite time average
$\bar{x}$, may appear.

The same distinction remains meaningful in the chaotic regime. A
banded-chaotic attractor may be globally invariant under $\mathcal{S}$, in
which case the coherent part of its spectrum obeys the same selection rule,
or it may occur as one member of a symmetry-related pair. In the latter case
$\bar{x}$ acquires opposite signs on the two attractors. Their merger at the
crisis discussed in Sec.~\ref{sec:loss} of the main text restores the symmetry and
drives $\bar{x}$ back to zero.

\section{Averaging theory for the $1{:}3$ resonance}

Here we provide the averaging calculation underlying the resonance tongue in
Fig.~\ref{fig:3}. We consider the Duffing equation in the weakly nonlinear
single-well regime,
\begin{equation}
\ddot{x}+\gamma\dot{x}+\omega_0^2x+\beta x^3
=
F\cos(\omega_d t),
\end{equation}
where the damping is small compared with the natural frequency,
$\gamma\ll\omega_0$, and the nonlinear restoring force remains a small
correction to the harmonic one over the relevant range of motion,
$|\beta|x^2\ll\omega_0^2$. We furthermore tune the drive close to the
$1{:}3$ resonance,
\begin{equation}
\left|\frac{\omega_d}{3}-\omega_0\right|\ll\omega_0 .
\end{equation}
It is convenient to introduce
\begin{equation}
\Omega\equiv\frac{\omega_d}{3}.
\end{equation}
Following the standard treatment of subharmonic resonances~\cite{NayfehMook1979},
we separate the response into a slowly varying subharmonic component at
$\Omega$ and the directly forced component at $\omega_d=3\Omega$,
\begin{equation}
x(t)
\simeq
a(t)\cos\psi
+
b\cos(\omega_d t-\delta),
\qquad
\psi=\Omega t+\theta(t).
\label{eq:SMansatz}
\end{equation}
To leading order, the forced component is given by the linear response,
\begin{equation}
b=
\frac{F}{
\sqrt{
(\omega_0^2-\omega_d^2)^2+\gamma^2\omega_d^2
}},
\label{eq:SMb}
\end{equation}
with phase lag satisfying
\begin{equation}
\tan\delta=
\frac{\gamma\omega_d}{\omega_0^2-\omega_d^2}.
\end{equation}

We now retain only terms resonant with the slow component at frequency
$\Omega$. The part of the cubic nonlinearity oscillating at this frequency is
\begin{equation}
\left[x^3\right]_{\Omega}
=
\left(
\frac{3}{4}a^3+\frac{3}{2}ab^2
\right)\cos\psi
+
\frac{3}{4}a^2b\cos(\psi-\Psi),
\label{eq:SMcubic}
\end{equation}
where
\begin{equation}
\Psi\equiv\delta+3\theta .
\end{equation}
The last term in Eq.~\eqref{eq:SMcubic} is the resonant mixing responsible
for the $1{:}3$ subharmonic response.

Projecting the equation of motion onto $\cos\psi$ and $\sin\psi$, and
neglecting second derivatives of the slowly varying amplitude and phase,
gives
\begin{align}
\dot a
&=
-\frac{\gamma}{2}a
+
\frac{3\beta}{8\Omega}a^2b\sin\Psi ,
\label{eq:SMslowA}
\\
\dot\theta
&=
\frac{1}{2\Omega}
\left[
\omega_0^2-\Omega^2
+\frac{3\beta}{4}a^2
+\frac{3\beta}{2}b^2
+\frac{3\beta}{4}ab\cos\Psi
\right].
\label{eq:SMslowTheta}
\end{align}
A stationary period-three response corresponds to
$\dot a=\dot\Psi=0$, i.e., $\dot a=\dot\theta=0$. For $a\neq0$,
Eqs.~\eqref{eq:SMslowA} and \eqref{eq:SMslowTheta} imply
\begin{align}
\gamma\Omega
&=
\frac{3\beta}{4}ab\sin\Psi,
\\
\omega_0^2-\Omega^2
+\frac{3\beta}{2}b^2
+\frac{3\beta}{4}a^2
&=
-\frac{3\beta}{4}ab\cos\Psi .
\end{align}
Eliminating the phase $\Psi$ gives
\begin{equation}
\left[
\omega_0^2-\Omega^2
+\frac{3\beta}{2}b^2
+\frac{3\beta}{4}a^2
\right]^2
+
(\gamma\Omega)^2
=
\left(\frac{3\beta}{4}\right)^2a^2b^2 ,
\label{eq:SMamplitude}
\end{equation}
which is Eq.~\eqref{eq:averaging} of the main text.

The saddle-node boundaries follow directly from
Eq.~\eqref{eq:SMamplitude}. Defining
\begin{equation}
u=a^2,\qquad
B=b^2,\qquad
\Delta=\omega_0^2-\Omega^2,\qquad
\Gamma=\gamma\Omega,
\qquad
c=\frac{3\beta}{4},
\end{equation}
the stationary amplitude equation becomes
\begin{equation}
\left(\Delta+2cB+cu\right)^2+\Gamma^2=c^2uB .
\label{eq:SMquadratic}
\end{equation}
This is a quadratic equation for $u$. At a saddle node its two roots
coalesce, so its discriminant vanishes. This gives
\begin{equation}
7c^2B^2+4c\Delta B+4\Gamma^2=0,
\end{equation}
and hence
\begin{equation}
b_\pm^2
=
\frac{8}{21\beta}
\left[
\Omega^2-\omega_0^2
\pm
\sqrt{
(\omega_0^2-\Omega^2)^2-7\gamma^2\Omega^2
}
\right].
\label{eq:SMbpm}
\end{equation}
Using Eq.~\eqref{eq:SMb} and $\Omega=\omega_d/3$, we finally obtain
\begin{align}
F_\pm(\omega_d)
=
\sqrt{
(\omega_0^2-\omega_d^2)^2
+\gamma^2\omega_d^2
}
\left[
\frac{8}{21\beta}
\left(
\left(\frac{\omega_d}{3}\right)^2-\omega_0^2
\pm
\sqrt{
\left[
\omega_0^2-\left(\frac{\omega_d}{3}\right)^2
\right]^2
-\frac{7\gamma^2\omega_d^2}{9}
}
\right)
\right]^{1/2},
\label{eq:SMtongue}
\end{align}
which gives the two boundaries plotted in Fig.~\ref{fig:3}. For $\beta>0$, real
solutions require
\begin{equation}
\omega_d>3\omega_0,
\qquad
\left[
\omega_0^2-\left(\frac{\omega_d}{3}\right)^2
\right]^2
\ge
\frac{7\gamma^2\omega_d^2}{9}.
\end{equation}
The averaged equations also clarify how the period-three response appears.
The harmonic solution, corresponding to $a=0$, remains stable for small
perturbations, since $\dot a\simeq-\gamma a/2$. The period-three response
therefore does not emerge continuously from this state. Instead, at the
boundary of the resonance tongue, Eq.~\eqref{eq:SMquadratic} develops two
nonzero solutions at a finite value of $a$. This is a saddle-node
bifurcation: a new period-three branch appears together with a second,
unstable branch.

Importantly, the original period-one response can remain stable at the same
parameters. The system can therefore support both a harmonic and a
period-three response, and which one is reached depends on the initial
condition. This explains the bistable region inside the resonance tongue
shown in Fig.~\ref{fig:3}.

\section{Feigenbaum hierarchy and numerical consistency checks}
\label{sec:SMfeig}

Here we provide the numerical data underlying the Feigenbaum analysis of
Fig.~\ref{fig:4} in the main text. All quantities reported in this section are obtained
from the same continuation, band-merging, and Lyapunov data used for that
figure, so that no independent estimate of the accumulation point or scaling
exponents is introduced in the Supplemental Material.

For the Duffing parameters
\begin{equation}
\alpha=-1,\qquad
\beta=1,\qquad
\gamma=0.2,\qquad
F=0.74,
\end{equation}
the asymmetric period-three branch undergoes the sequence of flip
bifurcations
\begin{equation}
3\rightarrow6\rightarrow12\rightarrow24\rightarrow48
\rightarrow96\rightarrow\cdots .
\end{equation}
The first five resolved flip points are listed in the upper part of
Table~\ref{tab:SMfeig}. Extrapolating the sequence gives the accumulation
point
\begin{equation}
\omega_c = 1.4349837(2),
\label{eq:SMomegac}
\end{equation}
where the uncertainty reflects the sensitivity of the extrapolation to the
last resolved levels. In the main text we quote the rounded value
$\omega_c\simeq1.434984$.

For a
$3\cdot2^k\rightarrow3\cdot2^{k-1}$ merger at $\omega_k$, we define
\begin{equation}
d_k=\omega_c-\omega_k.
\label{eq:SMdk}
\end{equation}
The merger locations, their numerical brackets, and the maximal Lyapunov
exponent evaluated at the same parameter values are listed in the lower part
of Table~\ref{tab:SMfeig}.

\begin{table}[t]
\centering
\caption{
Numerical hierarchy underlying the Feigenbaum analysis of Fig.~\ref{fig:4} in the
main text. The upper block lists the resolved flip bifurcations of the
asymmetric period-three branch. The lower block lists the independently
determined band-merging points $\omega_k$, the corresponding distances
$d_k=\omega_c-\omega_k$ with $\omega_c=1.4349837$, and the maximal
Lyapunov exponent evaluated at the same parameter values. Parentheses denote
the numerical uncertainty in the last quoted digits.
}
\label{tab:SMfeig}
\begin{ruledtabular}
\begin{tabular}{cccc}
event & $\omega$ & $d_k$ & $\lambda_{\max}$ \\
\hline
$3\rightarrow6$
& $1.43896423954$
& -- & -- \\

$6\rightarrow12$
& $1.43599029055$
& -- & -- \\

$12\rightarrow24$
& $1.43521242350$
& -- & -- \\

$24\rightarrow48$
& $1.43503355068$
& -- & -- \\

$48\rightarrow96$
& $1.43499445881$
& -- & -- \\
\hline
$6\rightarrow3$
& $1.433982319099$
& $1.0013809(35)\times10^{-3}$
& $0.0253751(174)$ \\

$12\rightarrow6$
& $1.434773156215$
& $2.1054379(73)\times10^{-4}$
& $0.0129936(109)$ \\

$24\rightarrow12$
& $1.434939744321$
& $4.395568(16)\times10^{-5}$
& $0.00650090(801)$ \\

$48\rightarrow24$
& $1.434974349367$
& $9.35063(3)\times10^{-6}$
& $0.00324764(541)$ \\

$96\rightarrow48$
& $1.434981741834$
& $1.958166(14)\times10^{-6}$
& $0.00162848(502)$
\end{tabular}
\end{ruledtabular}
\end{table}

The geometric scaling of the merger hierarchy can be tested directly from
the distances $d_k$. For the four resolved ratios we obtain
\begin{equation}
\ln\!\left(\frac{d_k}{d_{k+1}}\right)
=
1.5594,\;
1.5665,\;
1.5477,\;
1.5634 .
\label{eq:SMmergerspacings}
\end{equation}
Their mean and standard deviation are
\begin{equation}
\left\langle
\ln\!\left(\frac{d_k}{d_{k+1}}\right)
\right\rangle
=
1.559\pm0.008,
\label{eq:SMmergerspacingmean}
\end{equation}
in close agreement with the Feigenbaum prediction
\begin{equation}
\ln\delta_F = 1.54099,
\qquad
\delta_F\simeq4.6692016.
\end{equation}
The nearly constant logarithmic spacing also shows that the merging sequence
accumulates at the same $\omega_c$ as the flip sequence.

An independent and essentially fit-free test follows from the maximal
Lyapunov exponent. Huberman--Rudnick scaling predicts
\begin{equation}
\lambda_{\max}(d)
\propto d^{\nu_{\rm HR}},
\qquad
\nu_{\rm HR}
=
\frac{\ln2}{\ln\delta_F}
\simeq0.4498.
\label{eq:SMHR}
\end{equation}
Combining this relation with
$d_k/d_{k+1}\rightarrow\delta_F$ gives
\begin{equation}
\frac{\lambda_k}{\lambda_{k+1}}
\longrightarrow
\delta_F^{\nu_{\rm HR}}
=
2,
\qquad
\lambda_k\equiv\lambda_{\max}(\omega_k).
\label{eq:SMHRratio}
\end{equation}
From the values in Table~\ref{tab:SMfeig} we find
\begin{equation}
\frac{\lambda_k}{\lambda_{k+1}}
=
1.953,\;
1.999,\;
2.002,\;
1.994,
\label{eq:SMlambdaratios}
\end{equation}
in very close agreement with the parameter-free factor-of-two prediction.

The dense Lyapunov scan shown in Fig.~\ref{fig:4}(b) of the main text provides the
complementary continuous test of Eq.~\eqref{eq:SMHR}. Points belonging to
periodic windows are excluded from that scaling analysis. With the
log-period fixed to the Feigenbaum value
$\Delta_{\log}=\ln\delta_F$, the fit gives
$\nu=0.445\pm0.003$ (statistical error), with values in the range
approximately $0.44$--$0.45$ under variations of the fitting window. The
merger-point ratios in Eq.~\eqref{eq:SMlambdaratios} therefore provide the
more direct parameter-free test of the discrete Feigenbaum hierarchy.

Taken together, the convergence of the flip sequence, the geometric scaling
of the independently located band-merging points, and the factor-of-two
scaling of the Lyapunov exponent consistently identify the onset of banded
chaos with the Feigenbaum universality class.

\section{Band-merging crisis in the logistic map}

Here we provide additional diagnostics for the $6\rightarrow3$ reorganization
of the banded-chaotic attractor discussed in Fig.~\ref{fig:5}(a) of the main text.
The transition occurs at
\begin{equation}
r_c\simeq3.8511284,
\end{equation}
and corresponds to the pairwise merging of the six cyclic sectors into three.

\begin{figure}
    \centering
    \includegraphics[width=\linewidth]{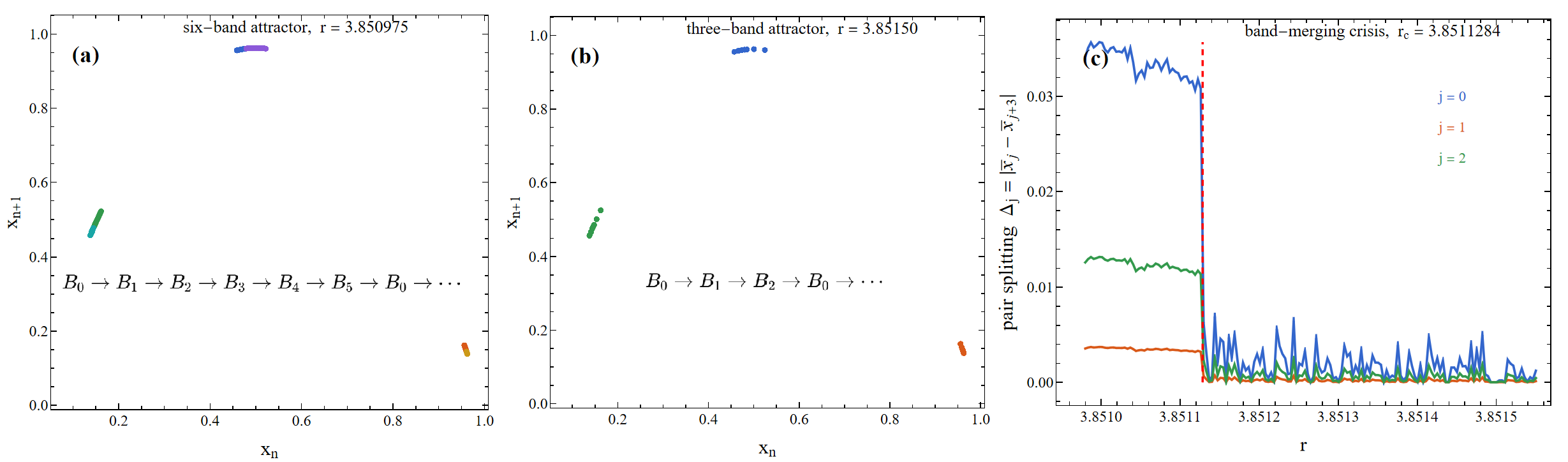}
    \caption{
\textbf{Pairwise merging of the six-band chaotic attractor in the logistic
map.}
\textbf{(a)} Six-band chaotic attractor at $r=3.850975$, below the
band-merging crisis. Colors distinguish the six cyclic classes
$n\bmod 6$.
\textbf{(b)} Three-band chaotic attractor at $r=3.85150$, above the crisis.
The former pairs of cyclic sectors have merged while the underlying
threefold organization remains.
\textbf{(c)} Pair splittings
$\Delta_j=|\bar{x}_j-\bar{x}_{j+3}|$, $j=0,1,2$, where
$\bar{x}_j=\langle x_{6\ell+j}\rangle_\ell$ is the conditional center of
the $j$th cyclic sector. All three splittings collapse at the
band-merging crisis $r_c=3.8511284$ (red dashed line), demonstrating the
pairwise identification
$B_j\leftrightarrow B_{j+3}$.
}
\label{fig:S2}
\end{figure}
To resolve this process geometrically, we define the conditional center of
each of the six phase classes,
\begin{equation}
\bar{x}_j
=
\left\langle x_{6\ell+j}\right\rangle_{\ell},
\qquad
j=0,\ldots,5,
\end{equation}
and the three pair splittings
\begin{equation}
\Delta_j
=
\left|
\bar{x}_j-\bar{x}_{j+3}
\right|,
\qquad
j=0,1,2.
\label{eq:SMpairSplitting}
\end{equation}
The geometric order parameter introduced in the main text is therefore simply
\begin{equation}
D_{63}^{\rm raw}
=
\frac{1}{3}
\sum_{j=0}^{2}\Delta_j .
\label{eq:SMD63}
\end{equation}

Figures~\ref{fig:S2}(a) and \ref{fig:S2}(b) show representative attractors
on the two sides of the transition. Below the crisis, the six phase classes
are distinct, although they are naturally organized into three nearby pairs.
Above the crisis, the members of each pair become dynamically
indistinguishable, leaving a three-band chaotic attractor. This pairwise
collapse is shown directly in Fig.~\ref{fig:S2}(c): all three
$\Delta_j$ remain finite on the six-band side and drop rapidly toward zero
at $r_c$.

For a representative six-band state at $r=3.850975$ we find
\begin{equation}
A_{1/6}\simeq7.2\times10^{-3},
\qquad
A_{1/3}\simeq0.236 ,
\end{equation}
showing that both the sixfold and the underlying threefold coherent
components are present below the crisis. Above the band-merging point, the $1/6$ component and the geometric
pair asymmetry vanish in the infinite-time limit. At finite observation
time both quantities retain small residual values that decrease systematically
with increasing trajectory length; this is the finite-time counterpart of
the spectral peak broadening at chaotic band mergings
~\cite{Brown1986,Romeiras1988}. For the trajectories used here these
residuals are below $10^{-5}$. By contrast, the $1/3$ component remains finite,
$A_{1/3}\simeq0.236$, confirming that the merger removes only the finer
sixfold temporal organization while preserving the underlying threefold
skeleton.
\section{Transverse instability of banded chaos}

Here we provide the analytical interpretation of the continuous
$3$-cyclic-to-$6$-cyclic reorganization studied in Fig.~\ref{fig:5}(b). The model is
\begin{align}
x_{n+1}
&=
r x_n(1-x_n),
\nonumber\\
y_{n+1}
&=
-\left[
1+\mu+c\left(x_n-\frac{1}{2}\right)
\right]y_n
+
g y_n^3 ,
\label{eq:SMtransverseMap}
\end{align}
with the logistic sector chosen inside its three-band chaotic regime. The
manifold $y=0$ is invariant and supports the original three-cyclic chaotic
attractor.

To determine its transverse stability, we linearize
Eq.~\eqref{eq:SMtransverseMap} about $y=0$,
\begin{equation}
\delta y_{n+1}
=
-a_n\,\delta y_n,
\qquad
a_n
=
1+\mu+c\left(x_n-\frac{1}{2}\right).
\label{eq:SMtransverseLinear}
\end{equation}
The transverse Lyapunov exponent is therefore
\begin{equation}
\lambda_\perp(\mu)
=
\lim_{N\rightarrow\infty}
\frac{1}{N}
\sum_{n=0}^{N-1}
\ln|a_n|
=
\left\langle
\ln\left|
1+\mu+c\left(x_n-\frac{1}{2}\right)
\right|
\right\rangle ,
\label{eq:SMlambdaPerp}
\end{equation}
where the average is taken over the natural measure of the three-band
attractor. The invariant manifold is stable for $\lambda_\perp<0$ and loses
transverse stability when
\begin{equation}
\lambda_\perp(\mu_c)=0.
\end{equation}
For the parameters used in Fig.~\ref{fig:5},
\begin{equation}
r=3.853,
\qquad
c=0.5,
\qquad
g=1,
\end{equation}
the state $y_n=0$ loses stability at
$\mu_c\simeq-1.85\times10^{-3}$, obtained from the condition
$\lambda_\perp(\mu_c)=0$.

Close to this point, the average growth rate varies approximately linearly,
\begin{equation}
\lambda_\perp(\mu)\propto \mu-\mu_c .
\end{equation}
For $\mu>\mu_c$, a small $y_n$ therefore tends to grow. This growth cannot
continue indefinitely because the cubic term in
Eq.~\eqref{eq:SMtransverseMap} becomes increasingly important at larger
$|y_n|$ and limits the amplitude. Balancing the growth close to threshold
with this nonlinear term gives
\begin{equation}
(\mu-\mu_c)|y|\sim g|y|^3 ,
\end{equation}
and hence
\begin{equation}
|y|\propto(\mu-\mu_c)^{1/2}.
\label{eq:SMsqrtTransverse}
\end{equation}
This explains the square-root onset observed in Fig.~\ref{fig:5}(b).

The appearance of the new $1/6$ component can also be understood directly.
We measure
\begin{equation}
O_n=x_n(1+y_n).
\label{eq:SMobservable}
\end{equation}
Below the transition, $y_n=0$, so the temporal structure is simply that of
the three-band chaotic logistic map, with fundamental reduced frequency
$q=1/3$. Above the transition, the minus sign in
Eq.~\eqref{eq:SMtransverseMap} makes $y_n$ alternate approximately from one
iteration to the next,
\begin{equation}
y_n\simeq(-1)^n |y_n|,
\end{equation}
corresponding to a period-two component, $q=1/2$.

The product $x_ny_n$ combines this period-two oscillation with the original
three-cycle of $x_n$. Their difference frequency is
\begin{equation}
\frac{1}{2}-\frac{1}{3}=\frac{1}{6},
\end{equation}
which produces the new coherent $1/6$ component. Since its amplitude is
proportional to $|y|$ close to the transition,
\begin{equation}
|\Phi_{1/6}|
\propto
|y|
\propto
(\mu-\mu_c)^{1/2}.
\label{eq:SMphi16Scaling}
\end{equation}

The strong $1/2$ peak has an even simpler origin. Since $x_n$ has a nonzero
average value, the observable $O_n=x_n(1+y_n)$ contains a contribution
proportional directly to $y_n$, and therefore inherits its period-two
oscillation.

Very close to $\mu_c$, the chaotic fluctuations of $x_n$ make the growth of
$y_n$ irregular: long intervals with very small amplitude are interrupted by
bursts. This intermittent behavior modifies the simple square-root scaling
extremely close to the transition~\cite{PlattSpiegelTresser1993,OttSommerer1994}.
For the parameters considered here, these deviations become appreciable only
below the range shown in Fig.~\ref{fig:5}(b).

The important point is therefore simple: the dynamics of $x_n$ remains
chaotic throughout the transition, while the onset of a period-two component
in $y_n$ changes the temporal organization from three to six cyclic sectors.

\section{Attractor-merging crisis in the Duffing oscillator}

Here we provide additional dynamical evidence for the final loss of
time-semicrystalline order discussed in Fig.~\ref{fig:6} of the main text. For
\begin{equation}
\alpha=-1,\qquad
\beta=1,\qquad
\gamma=0.30,\qquad
F=0.58,
\end{equation}
the transition occurs in the narrow interval
\begin{equation}
1.39398<\omega^\star<1.39399.
\end{equation}
Immediately below $\omega^\star$, the dynamics consists of two distinct
chaotic attractors related by the spatiotemporal $\mathbb{Z}_2$ symmetry.
Each attractor retains a three-band cyclic organization. Because their basins
are interleaved close to the crisis, different initial conditions were used
to select representative trajectories on the two branches.

\begin{figure}
    \centering
    \includegraphics[width=\linewidth]{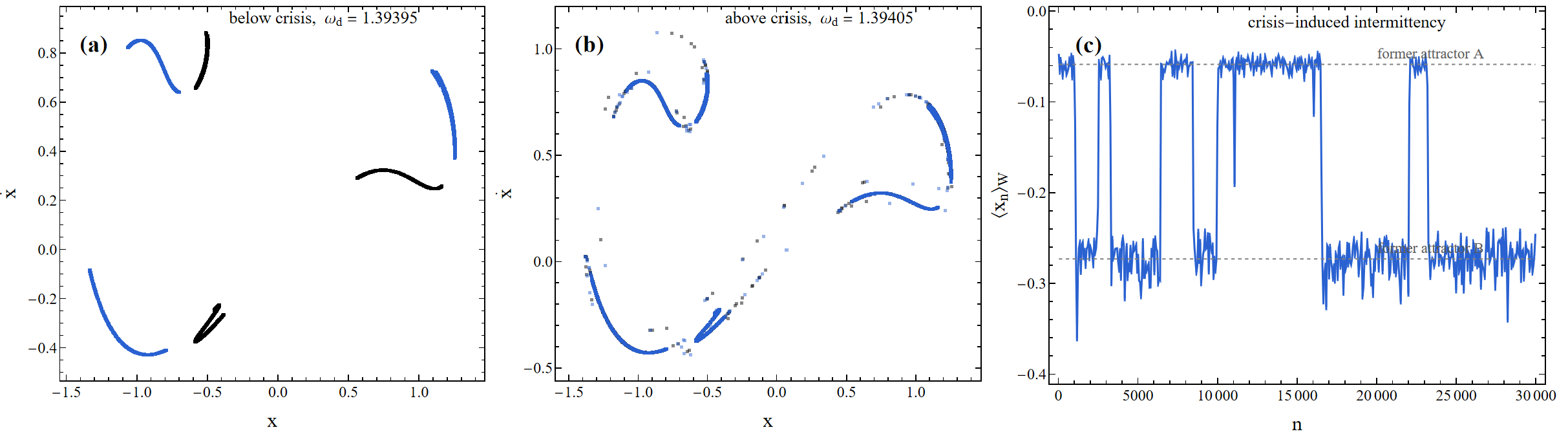}
    \caption{
\textbf{Attractor-merging crisis and crisis-induced intermittency in the
driven Duffing oscillator.}
The parameters are $\alpha=-1$, $\beta=1$, $\gamma=0.30$, and $F=0.58$.
\textbf{(a)} Immediately below the crisis, at
$\omega_d=1.39395$, two distinct symmetry-related three-band chaotic
attractors coexist. The black and blue points correspond to trajectories
initialized in the two different basins.
\textbf{(b)} Above the crisis, at $\omega_d=1.39405$, the two attractors
have merged into a single enlarged chaotic set; trajectories starting from
the same two initial conditions now explore the same attractor.
\textbf{(c)} Crisis-induced intermittency slightly above the transition,
at $\omega_d=1.39400$: block-averaged stroboscopic displacement
$\langle x_n\rangle_W$ with $W=60$ driving periods. Dashed lines mark the
stroboscopic levels of the two former attractors; the trajectory switches
intermittently between them.
}
\label{fig:S3}
\end{figure}
Figure~\ref{fig:S3}(a) shows the corresponding stroboscopic Poincar\'e
sections at $\omega_d=1.39395$. The two chaotic sets are disconnected and
remain dynamically inaccessible from one another. At the crisis they collide
with the intervening unstable structure and merge. Above the transition,
Fig.~\ref{fig:S3}(b), trajectories initialized in the two former basins
populate the same enlarged chaotic attractor. The crisis therefore removes
the dynamical separation between the two symmetry-related chaotic states.

A characteristic signature of this merger appears in the dynamics
immediately above the crisis. To visualize it, we introduce the
block-averaged stroboscopic displacement
\begin{equation}
\langle x_n\rangle_W
=
\frac{1}{W}
\sum_{j=0}^{W-1} x_{n+j},
\qquad
W=60,
\label{eq:SMblockaverage}
\end{equation}
where $x_n=x(nT_d)$. This quantity is used only as a diagnostic of the
stroboscopic dynamics and should not be confused with the continuous-time
symmetry-breaking order parameter
$\bar{x}=\langle x(t)\rangle_t$ employed in the main text.

As shown in Fig.~\ref{fig:S3}(c), slightly above the crisis the merged
trajectory still spends long intervals close to the remnants of the two
pre-crisis attractors. These intervals are interrupted by rapid transitions
between the two regions of phase space. This intermittent switching is the
standard dynamical signature of a merging crisis: immediately after the two
formerly invariant chaotic sets become connected, transport between them
remains rare on short time scales.

The final transition can therefore be summarized as
\begin{equation}
\text{two symmetry-related banded-chaotic attractors}
\;\longrightarrow\;
\text{one merged chaotic attractor}.
\end{equation}
The disappearance of the coherent $1/3$ component and the restoration of the
$\mathbb{Z}_2$ symmetry shown in Fig.~\ref{fig:6} of the main text are thus accompanied
by the expected geometrical and dynamical signatures of an
attractor-merging crisis.

\end{document}